\documentclass[longbibliography,reprint,onecolumn,showpacs,superscriptaddress]{revtex4-2}%

\usepackage{multirow}
\usepackage{mathrsfs,amsmath}
\usepackage{amsfonts}
\usepackage{amssymb}
\usepackage{amsthm}
\usepackage{bm}
\usepackage{xparse}
\usepackage{graphicx}
\usepackage[dvipsnames]{xcolor}
\usepackage{appendix}
\usepackage{enumitem}
\usepackage{dsfont}
\usepackage{colonequals}
\usepackage{bbold}

\usepackage[normalem]{ulem}
\usepackage{caption}
\usepackage{subcaption}
\newcommand{\ket}[1]{| {#1} \rangle} % for Dirac kets
\newcommand{\kket}[1]{| {#1} \rangle\rangle} % for doublekets
\newcommand{\bra}[1]{\langle {#1} |} % for Dirac bras
\newcommand{\braket}[2]{\langle {#1} \vphantom{#2} | {#2} \vphantom{#1} \rangle} % for Dirac brackets
\newcommand{\ketbra}[2]{| {#1} \vphantom{#2} \rangle\langle {#2} \vphantom{#1} |} % for Dirac brackets
\newcommand{\proj}[1]{|#1\rangle\langle#1|}

\DeclareMathOperator*{\ox}{\otimes}		%for tensor products
\DeclareMathOperator{\HH}{\mathcal{H}}
\DeclareMathOperator{\LL}{\mathcal{L}}
\DeclareMathOperator{\UU}{\mathcal{U}}

\DeclareMathOperator{\id}{\mathbb{1}}
\DeclareMathOperator{\rrangle}{\rangle\rangle}

\begin{document}

\title{Noncausal Page-Wootters circuits}

\author{Veronika Baumann}
\thanks{These two authors contributed equally to this paper.}
\affiliation{Faculty of Physics, University of Vienna, Boltzmanngasse 5, 1090 Vienna, Austria}
\affiliation{Institute for Quantum Optics and Quantum Information (IQOQI), Austrian Academy of Sciences, Boltzmanngasse 3, 1090 Vienna, Austria}
\affiliation{Faculty of Informatics, Universit\`a della Svizzera italiana, Via G. Buffi 13, CH-6900 Lugano, Switzerland}

\author{Marius Krumm}
\thanks{These two authors contributed equally to this paper.}
\affiliation{Faculty of Physics, University of Vienna, Boltzmanngasse 5, 1090 Vienna, Austria}
\affiliation{Institute for Quantum Optics and Quantum Information (IQOQI), Austrian Academy of Sciences, Boltzmanngasse 3, 1090 Vienna, Austria}

\author{Philippe Allard Gu\'{e}rin}
%\thanks{These three authors contributed equally to this work.}
\affiliation{Faculty of Physics, University of Vienna, Boltzmanngasse 5, 1090 Vienna, Austria}
\affiliation{Institute for Quantum Optics and Quantum Information (IQOQI), Austrian Academy of Sciences, Boltzmanngasse 3, 1090 Vienna, Austria}
\affiliation{Perimeter Institute for Theoretical Physics, 31 Caroline St. N, Waterloo, Ontario, N2L 2Y5, Canada}

\author{\v{C}aslav Brukner}
\affiliation{Faculty of Physics, University of Vienna, Boltzmanngasse 5, 1090 Vienna, Austria}
\affiliation{Institute for Quantum Optics and Quantum Information (IQOQI), Austrian Academy of Sciences, Boltzmanngasse 3, 1090 Vienna, Austria}

\date{\today}
%
%========================================================================%
\begin{abstract}
%The Page-Wootters formalism is a popular approach towards a quantum theory of gravity because it treats space and time on equal footing.
One of the most fundamental open problems in physics is the unification of general relativity and quantum theory to a theory of quantum gravity. An aspect that might become relevant in such a theory is that the dynamical nature of causal structure present in general relativity displays quantum uncertainty. This may lead to a phenomenon known as indefinite or quantum causal structure, as captured by the process matrix formalism. Due to the generality of that framework, however, for many process matrices there is no clear physical interpretation. 
A popular approach towards a quantum theory of gravity is the Page-Wootters formalism, which associates to time a Hilbert space structure similar to  spatial position. By explicitly introducing a quantum clock, it allows to describe time-evolution of systems via correlations between this clock and said systems encoded in history states.
In this paper we combine the process matrix framework with a generalization of the Page-Wootters formalism in which one considers several agents, each with their own discrete quantum clock. We describe how to extract process matrices from scenarios involving such agents with quantum clocks, and analyze their properties. The description via a history state with multiple clocks imposes constraints on the implementation of process matrices and on the perspectives of the agents as described via causal reference frames. While it allows for scenarios where different definite causal orders are coherently controlled, we explain why certain non-causal processes might not be implementable within this setting. \\
\end{abstract}
%========================================================================%
\maketitle
\setcounter{page}{1}

%%%%%%%%%%%%%%%%%%%%%%%%%%%%%%%%%%%%%%%%%%%%%%%%%%%%%%%
\section{Introduction}
\label{Introduction}

%One of the most fundamental open problems in physics is the unification of general relativity and quantum theory to a theory of quantum gravity. 

Indefinite causal structure is an extension of the usual notion of causal structure that is expected to become relevant in quantum gravity:  In general relativity, causal structure is dynamical instead of fixed and attributing quantum properties~\cite{Hardy2005, Hardy2007, Hardy2009, Feynman1957, zych2019bell, Rovelli2004, Kiefer2012, Butterfield1999} would imply the existence of exotic causal structures, like superpositions of space-times and superpositions of the order of events. The process matrix framework~\cite{OCB,araujo2017purification} was invented to systematically describe such indefinite or quantum causal structures. However, many processes that arise in this framework have no clear physical interpretation and it is not known which of them are realizable in nature. It has, therefore, been suggested that only processes, which reversibly map a well defined causal past to a well defined causal future with possibly indefinite causal order in between, are physical~ \cite{araujo2017purification}. 
For such processes, it has been shown that one can always find a causal reference frame that represents the perspective of an agent within the causal structure~\cite{guerin2018agent}. While the agent's event is local in their causal frame of reference, the events of other agents may be ``smeared'' over the causal past and future of the event. Still the question which processes are realizable in nature remains open.
Some non-causal processes can violate device-independent causal inequalities~\cite{branciard2015simplest, baumeler2014maximal,brukner2015bounding}, although no physical interpretation for such processes are known. Other processes, for example the so-called quantum switch~\cite{chiribella2013quantum,procopio2015experimental,rubino2017experimental}, where the order of operations is controlled by a quantum system, cannot violate causal inequalities but exhibit indefinite causal order that can be identified by causal witnesses~\cite{araujo2015witnessing}. Moreover, it is possible to experimentally implement such coherent quantum control of causal order~\cite{procopio2015experimental,rubino2017experimental, wei2019experimental, goswami2018communicating, Taddei2020, Rubino2017, Goswami2018, Guo2020}. Overall, however, many mathematically consistent process matrices are believed to be unphysical, yet there are only very few reasonable postulates, like~\cite{araujo2017purification}, that actually allow to rule out some processes as unphysical.\\

A crucial obstacle in finding a complete theory of quantum gravity is caused by the different role of time in general relativity and quantum theory. As an approach to bridge this conceptual gap, one can use a timeless formalism~\cite{Page1983, Unruh1989, Rovelli1990, Reisenberger2002, Hellmann2006, Giovannetti2015, Hoehn2018,hohnSwitchingInternalTimes2019a, Hoehn2019, Hoehn2020, Castro-Ruiz2020}, which we refer to as the Page-Wootters formalism in this paper. In this formalism, one also associates a Hilbert space with time, which can be interpreted as a quantum clock. One describes the physics of the extended system including the clock by using history states which are obtained via a Wheeler-DeWitt-like equation using a constraint operator. These history states encode dynamics as correlations between the main system and the quantum clock.
In Ref.~\cite{Castro-Ruiz2020}, the authors considered a generalized Page-Wootters approach using several clocks. The authors found that history states arising from solving a Hamiltonian constraint for gravitationally interacting clocks can give rise to indefinite causal order and studied the time evolution according to the perspectives of different clocks. In particular, they showed how the Page-Wootters formalism can recover the so-called gravitational quantum switch~\cite{zych2019bell}. Their approach works for important examples, but it is not clear in general which process (if any) is implemented by a given history state, or what is the set of non-causal processes that can be implemented within such a framework. 
%
%\textcolor{red}{
%\sout{Some non-causal processes can violate device-independent causal inequalities}~\cite{branciard2015simplest, baumeler2014maximal,brukner2015bounding}, \sout{although no physical interpretation for such processes are known. Other processes, for example the so-called quantum switch}~\cite{chiribella2013quantum,procopio2015experimental,rubino2017experimental}, \sout{where the order of operations is controlled by a quantum system, cannot violate causal inequalities but exhibit indefinite causal order that can be identified by causal witnesses}~\cite{araujo2015witnessing}. \sout{Moreover, it is possible to experimentally implement such coherent quantum control of causal order}~\cite{procopio2015experimental,rubino2017experimental, wei2019experimental, goswami2018communicating, Taddei2020, Rubino2017, Goswami2018, Guo2020}.}\\
%
The observations in~\cite{Castro-Ruiz2020} motivate us to investigate the relation between causal order and several quantum clocks in further detail, and develop a full framework that systematically combines the process matrix and the Page-Wootters formalism. A refined description of indefinite causal structure that explicitly models the agents' perceptions of time might help to discover physical scenarios that give rise to indefinite causal structure. In particular, a well-defined combination of these two approaches might lead to new consistency conditions that allow to rule out some process matrices as unphysical. \\

Even beyond the search for criteria for physical process matrices, the combination of process matrices with the Page-Wootters formalism is an interesting topic by itself. Both frameworks can be seen as approaches towards quantum gravity. However, so far they were mostly isolated and independent from each other, although they are not incompatible. The Page-Wootters formalism can be regarded as an independent formulation of quantum theory, similar to the path-integral formulation. It is an interesting question which processes can be naturally implemented within a particular formulation of quantum theory. It is known that any quantum circuit (i.e. a definite causal order) can be implemented within the Page-Wootters formalism as a Feynman's quantum computer and a single (i.e. global) quantum clock~\cite{feynman1985quantum, kitaev2002classical, Breuckmann2014, Caha2018}. Can the formalism also account for more general causal structures such as coherently controlled causal orders or even those that lead to the violation of causal inequalities?
To approach this question, in the present paper, we present a general definition of what it means for a history state to implement a pure process matrix~\cite{araujo2017purification}, for the case of finite dimensional systems and many clocks. 
We describe how to extract the agents' perspectives from the history states, which corresponds to a refinement of the concept of causal reference frames~\cite{guerin2018agent} that explicitly includes the quantum clocks. We show that arbitrary coherently controlled causal order can be extracted from our framework when different clocks tick at different rates (for example, due to time dilation effects). Moreover, we analyze the additional restrictions that the history states impose on the extracted process matrices. These restrictions might be regarded as reasons why some processes cannot be implemented in nature. While the Page-Wootters formalism with many clocks can enable the extraction of processes with definite causal order and quantum controlled causal order, it additionally provides insights into why some processes might not be realizable within the framework.
We consider discrete instead of continuous clocks because this allows us to express the perspectives of the agents using circuits. Our approach does not start by defining a constraint operator and solving it; instead we work directly at the level of history states. We develop a systematic framework that combines process matrices and Page-Wootters history states with several discrete clocks. We describe how to model scenarios where these clocks are associated with agents that are initially all part of a definite space-time causal structure. Then the agents might enter a "region" of quantum causal structure, where the global order of events is no longer well defined. At the end, however, all agents return to a definite causal structure. \\ %We show how to extract the process matrix from such scenarios and analyze the additional restrictions that the history states impose on the extracted process matrices. Such restrictions can be regarded as an argument against the implementability of those processes that are excluded by them. 
%We further describe how to extract the agents' perspectives form the history states, which corresponds to a refinement of the concept of causal reference frames~\cite{guerin2018agent} that explicitly includes the quantum clocks. We show how to describe arbitrary coherently controlled causal order in our framework.\\

The paper is structured as follows: We first recapitulate important aspects of the process matrix formalism (including causal reference frames) and the Page-Wootters formalisms in Sections~\ref{Process matrices and causal reference frames} and~\ref{The Page-Wootters formalism} before we motivate and introduce our framework, which combines these two approaches, in detail in Section~\ref{Process matrices within a timeless formalism}. In that context, we derive several mathematical properties of our setting, in particular restrictions on process matrices. In Section~\ref{Non-causal Page-Wootters circuits}, we first construct examples involving varying clock speeds and indefinite causal structure before we show how to implement arbitrary quantum-controlled causal order in our setting. At the end of Section~\ref{Non-causal Page-Wootters circuits} we discuss why another well-known, non-causal process might not be implementable in our framework. Finally, we discuss our findings in Section~\ref{Discussion}. 

%%%%%%%%%%%%%%%%%%%%%%%%%%%%%%%\ref%%%%%%%%%%%%%%%%%%%%%%%%

\section{Process matrices and causal reference frames}
\label{Process matrices and causal reference frames}
%In general relativity, the causal structure is dynamical rather than fixed in advance. In quantum gravity, such a dynamical causal structure would have quantum properties~\cite{Hardy2005, Hardy2007, Hardy2009, Feynman1957, Zych2019, Rovelli2004, Kiefer2012, Butterfield1999}. This would imply the existence of exotic causal structures, like superpositions of space-times and superpositions of the order of events. The process matrix formalism~\cite{OCB} was invented to systematically describe such causal structures. 
In this section we give a short introduction to the operational setting of the process matrix formalism and explain the parts of the framework that are important for the rest of the paper.\\ 

\begin{figure}[h]
\begin{subfigure}{0.44\textwidth}
\vspace{3 em}
\includegraphics[width=1.1\linewidth]{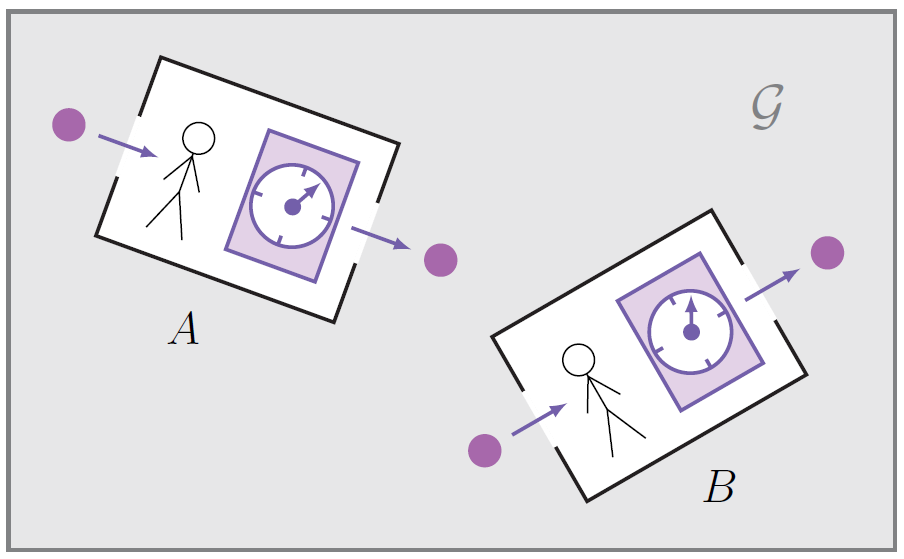} 
\vspace{1.5 em}
\caption{Basic setting}
\label{Fig:ProcessSetting}
\end{subfigure}
\hfill
\begin{subfigure}{0.44\textwidth}
\includegraphics[width=\linewidth]{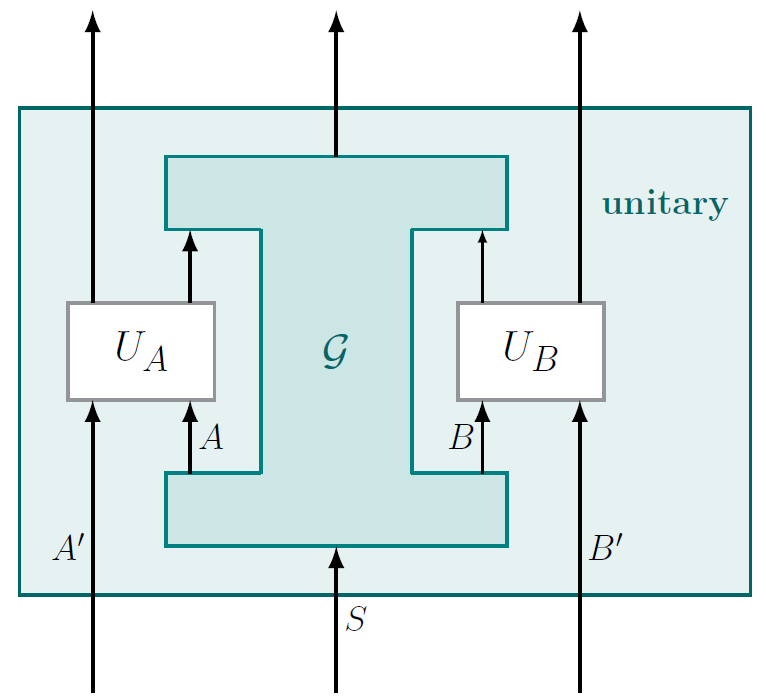} 
\caption{A pure process}
\label{Fig:Pure}
\end{subfigure}
\caption{ Example of bipartite processes:
The two agents (here called $A$ and $B$) are each situated in their own laboratory. Each agent obtains a system from the environment, acts on it with a quantum instrument and then sends it out again. While inside the laboratories the order of events is well defined, there need not be a well defined global ordering imposed by the environment. The outcome statistics of the operations performed by $A$ and $B$ is described by a process matrix $\mathcal G$, see (\ref{Fig:ProcessSetting}).
These quantum instruments of $A$ and $B$ can be represented as unitaries $U_A, U_B$ by introducing ancillary systems $A', B'$. A pure process $\mathcal G$ is a (multilinear) supermap that gives an induced unitary transformation on $S \otimes A' \otimes B'$ when the agents are applying unitary operations $U_A, U_B$, see (\ref{Fig:Pure}).
}
\label{Fig:ProcessBasics}
\end{figure}

The basic operational setting of the process matrix formalism concerns several agents (here $N$ of them), labeled $A_1 \dots A_N$, each of them inside their own (small) laboratory where the usual rules of quantum theory are valid. The outside ``environment'', which relates the various agents, is not assumed to be causally definite, for example it could be a superposition of space-time structures. During the protocol, each agent receives a quantum system from the ``environment'', applies a quantum instrument i.e. a probabilistic quantum channel (for example a measurement or a pure quantum channel) to that system and then sends it out again. This well defined local time ordering inside the laboratory can be thought of as being tracked by a clock associated with each agent, the bipartite case is depicted in Figure \ref{Fig:ProcessSetting}.
The process matrix $\mathcal G$ is the mathematical object that encodes the observed outcome probabilities for any choice of local quantum instruments. Process matrices describing definite causal order, i.e.  the global order of operations performed by different agents is well defined, are equivalent to higher order quantum maps or quantum combs ~\cite{Chiribella_2008,Perinotti_2017,CombsLong,CombsPRA,TulioSuperchannels}. In general, however, they allow scenarios with indefinite causal order, where no such global order exists, and are in that sense generalizations of quantum combs.  \\

%The basic operational setting of the process matrix formalism is pictured in Figure~\ref{Fig:ProcessSetting}. We consider several agents, each of them inside their own (small) lab. One assumes that within each lab, the usual rules of quantum theory are valid. The outside ``environment'', which relates the various agents together, is not assumed to be causally definite, for example it could be a superposition of space-time structures. During the protocol, each agent sees a quantum system enter their lab. The agent can apply a quantum instrument (for example a measurement or a quantum channel) to the system and then sends it out again. The process matrix is the mathematical object that encodes the observed outcome probabilities for any choice of local quantum instruments. %The crucial difference to the usual process matrix framework will be that we model the passage of time in the labs by the use of quantum clocks.

In \cite{araujo2017purification}, processes (called quantum superchannels in \cite{TulioSuperchannels}) are formalized as maps from a global past to a global future, that depend on the agents' operations.
In addition to the agent's systems, whose Hilbert spaces are labeled $A_1 \dots, A_N$, we introduce ancillary systems $A_1', \dots, A_N'$. The ancillary system can be used, for example, as a quantum memory recording a measurement outcome. Each agent is allowed to act with a quantum channel on their system and their ancilla, i.e. a completely positive trace-preserving map (CPTP) $\mathcal C_{A_j} : \mathcal L(A_j A'_j) \rightarrow \mathcal L(A_j A'_j)$. One assumes that the ancillas have trivial evolution except when the respective operations of the agents are applied. Then, a \emph{process} (or \emph{quantum superchannel}) is a multilinear map $\mathcal G$ that maps the agents' quantum channels to a quantum channel, while acting as the identity on all the ancillary systems (just as in Figure \ref{Fig:Pure}, but with the unitaries replaced by quantum channels). This map encodes the causal structure given by the environment. \\

In this paper, we only consider pure processes. 
Using Stinespring's dilation theorem~\cite{Stinespring,NielsenChuang} one can represent the quantum operations of the agents as unitaries $U_{A_1} \dots U_{A_N}$  acting on the respective system and ancilla that the agents obtain from the environment. The ancillas serve as both purifying systems and as memories recording the outcomes. We say that a process $\mathcal{G}$ is a pure process if it is a unitary preserving map, i.e. $\mathcal{G}(U_{A_1} \dots U_{A_N})$ is unitary for any unitaries $U_{A_1} \dots U_{A_N}$, while acting as the identity on the ancillary systems. The basic mathematical structure of a bipartite pure process is depicted in Figure~\ref{Fig:Pure}. \\

Since many non-causal process matrices lack a clear interpretation and it is not clear whether they are compatible with the known physical laws, it has been suggested that only purifiable processes, which means they can be obtained from pure processes, are physical~\cite{araujo2017purification}. Such processes can be regarded as reversible transformations from a well defined causal past to a well defined causal future, with indefinite causal order in between. We note that the quantum-switch is an example of a pure process, as shown explicitly in~\cite{araujo2017purification}, and it is physically realizable either in gravitational~\cite{Castro-Ruiz2020, zych2019bell} or optical setups~\cite{procopio2015experimental,rubino2017experimental, wei2019experimental, goswami2018communicating, Taddei2020, Rubino2017, Goswami2018, Guo2020}.\\

%For simplicity we consider the bipartite case, but the generalization to more agents is straightforward. By introducing ancillary systems $A', B'$ one can represent the quantum instruments of the agents as unitaries $U_A, U_B$  acting on the respective ancilla and the system $A,B$ that the agents obtain from the environment $\mathcal G$. The ancillas serve as purifying systems and as memories recording the outcomes. One assumes that the ancillas have trivial evolution except when $U_A, U_B$ are being applied. \textcolor{red}{The causal structure imposed by the environment is modeled by a map $\mathcal G(U_A,U_B)$ that maps an input $S$ from the global past to an output to the global future.}  \textcolor{red}{We say that $\mathcal{G}$ is a pure process if it is a multi-linear unitary preserving map, i.e. $\mathcal{G}(U_A, U_B)$ is unitary for any two unitaries $U_A, U_B$.} 
%We say that $\mathcal{G}$ is a pure process if the map $(U_A, U_B) \mapsto \mathcal{G}(U_A, U_B)$ is multi-linear and unitarity-preserving.\\

The notion of causal reference frames~\cite{guerin2018agent} was introduced as an equivalent description of the pure process matrix formalism. The causal reference frame represents the perspective of an agent inside a (possibly indefinite) causal structure. More concretely, one imagines the perspective of an agent, say $A_1$, as follows: The crucial moment for agent $A_1$ is when he or she applies unitary $U_{A_1}$. The evolution starting from the beginning of the protocol up to that moment is described by a unitary $\Pi_ {A_1}(U_{A_2} \dots U_{A_N})$, which is called the causal past of $A$ and can depend on the instruments of all other agents. Then $A_1$ enforces time evolution via $U_{A_1}$ on the input to his or her laboratory and the ancilla $A_1'$, while all other degrees of freedom evolve in an uncorrelated way. The evolution of these other degrees of freedom can be absorbed into $\Pi_{A_1}(U_{A_2} \dots U_{A_N})$ such that without loss of generality we can assume that during $A_1$'s time of action, evolution is given by $U_A \otimes \mathbb 1$. Afterwards the evolution up to the end of the protocol is described by a unitary $\Phi_{A_1}(U_{A_2} \dots U_{A_N})$, which is called the causal future of $A_1$. It can again depend on the instruments of all other agents. As shown in Ref.~\cite{guerin2018agent} all pure processes admit a decomposition in causal reference frames, i.e. if $\mathcal{G}$ is a pure process, then $\mathcal{G}$ can be written as
\begin{equation}
\mathcal{G}(U_{A_1} \dots U_{A_N}) = \Phi_{A_1}(U_{A_2} \dots U_{A_N}) \left( U_{A_1} \otimes \id \right) \Pi_{A_1}(U_{A_2} \dots U_{A_N}), \label{Equation:GfromCausalRefFrames}
\end{equation}
where $\Phi_{A_1}(U_{A_2} \dots U_{A_N}), \Pi_{A_1}(U_{A_2} \dots U_{A_N})$ are unitaries that depend on all unitaries other than $U_{A_1}$ and that describe the time-evolution according to $A_1$'s point of view. A similar decomposition exists from the point of view of all other agents. In the present paper we take a similar approach, but we make the addition of a localized quantum clock associated to each agent, and explain how the perspectives of various agents can arise from a perspective neutral history state as given by the Page-Wootters formalism.

 % Like in our framework, each agent has a specific moment when they act with their local instrument, described via purified unitaries $U_A$ ($U_B$, $U_C$). The change from the beginning of the experiment to the agent's moment of action is described by a unitary. This is interpreted as the past the agents see before they have to act. In our framework, this unitary would correspond $\UU_A(t_A^*-1,0)$. Similarly, there is an unitary that relates the moment after the action to the end of the experiment. This is interpreted as the future that the agents see after their actions. In our framework, this would correspond to $\UU_A(T,t_A^*)$. \textcolor{blue}{[We have to find a way to say what we want without notation that hasn't been introduced yet]} \\

%The crucial difference between our framework and \citep{guerin2018agent} is that we explicitly model the quantum clocks and explain how the agents perspectives arise from a perspective neutral history state. As we will show in this paper, this imposes further strong conditions on the past the agents see before acting and the future they see after acting.\\

%%%%%%%%%%%%%%%%%%%%%%%%%%%%%%%%%%%%%%%%%%%%%%%%%%%%%%%

\section{The Page-Wootters formalism}
\label{The Page-Wootters formalism}

In this section we give a brief general overview of the Page-Wootters formalism for continuous as well as discrete quantum clocks. A possible justification for considering quantum clocks with discrete Hilbert spaces comes from arguments involving the Bekenstein bound~\cite{Bekenstein1973} that Hilbert space is fundamentally finite-dimensional~\cite{Garay1994, Bao2017}. Also, it can be argued that all information that can ever be acquired via measurements is finite and that therefore on the fundamental level physics should be discrete as well and indeed, finite~\cite{Gisin2019}. Furthermore, we get significant technical simplifications due to the fact that for finite-dimensional Hilbert spaces, the physical Hilbert space is a subspace of the kinematical Hilbert space, while this is not the case for infinite dimensional~\cite{Marolf2000, Hoehn2019}.
Most importantly for our purpose, the assumption of finite dimensional clocks allows us to get a physical picture of indefinite causal structure in form of generalizations of quantum circuits.\\

%The Page-Wootters formalism~\cite{Page1983} is a way to treat space and time on equal footing in quantum mechanics. 
In addition to the usual system Hilbert space $\HH_S$ the Page-Wootters formalism introduces an additional Hilbert space $\HH_c$ associated with time that can be interpreted as an ideal quantum clock. In analogy to position in non-relativistic quantum mechanics $\HH_c$ can be chosen to be spanned by square integrable functions on the real line; informally it is common in physics to imagine this Hilbert space as $\HH_c= \mathrm{span}\{ \ket{t}\ | \ t \in \mathbb R \} $.
In analogy to the usual momentum operator, one can define an operator $\hat{p}_t$ as the generator of translations on $\HH_c$. In the time representation, i.e. $\braket{t}{\psi}$, it is given by $\hat{p}_t = -i \frac{\partial}{\partial t}$.
Let $\hat{H}_S$ be the Hamiltonian of the system and consider the constraint operator $\hat{C} := \hat{p}_t+ \hat{H}_S $. Let $\ket{\Psi \rangle}$ be a state on $\HH_c \otimes \HH_S$ that satisfies the Wheeler-DeWitt-like constraint equation $\hat{C} \ket{\Psi \rangle} = 0$. Such states $\kket{\Psi}$ are often called physical states.
Without worrying about normalizability, one can formally expand $\ket{\Psi \rangle}$ by using the time basis as
\begin{align}
	\ket{\Psi \rangle} = \int \mathrm{d} t\ \ket{t} \otimes \ket{\psi(t)}. \label{Equation:History}
\end{align}  
With this expansion it becomes clear why states $\kket{\Psi}$ are also called history states: For each time $t$, they encode a system state $\ket{\psi(t)}$ and an ordered time sequence $t_0<t_1<t_2$ corresponds to the history of the state given by $\ket{\psi(t_0)},\ket{\psi(t_1)}$ and $\ket{\psi(t_2)}$.
%The constraint equation  $\hat{C} \ket{\Psi \rangle} = 0$ looks like a Schr\"{o}dinger equation without (background) time, further earning $\ket{\Psi \rangle}$ the name \emph{time-less state}. 
Plugging the expansion Eq. (\ref{Equation:History}) into the constraint equation, one can show that the system state satisfies
\begin{align}
	i \frac{\partial}{\partial t} \ket{\psi(t)} = H_S \ket{\psi(t)},
\end{align}
which is the standard Schr\"{o}dinger equation. Therefore, this approach recovers the usual quantum formalism.
In general, solutions to the constraint equation can be obtained via an operator
\begin{align}
\hat{P} := \int_{\mathbb{R}} ds \, e^{-is\hat{C}}, 
\label{Pphys}
\end{align}
which gives a valid physical (or history) state, i.e. solution to the constrain equation $\hat{C}(\hat P \ket{\phi})=0$, when applied to arbitrary states $\ket{\phi}\in \mathcal{H}_c \otimes \mathcal{H}_S$. For this reason the operator $\hat P$ is sometimes called the physical projector \cite{dolby2004conditional}, although it is not a projector in the strict mathematical sense.
Moreover, $\bra{t_2}\hat P\ket{t_1}=\mathcal{U}(t_2,t_1)$ is a unitary operator on $\mathcal{H}_S$ and in case of there being no interaction term between clock and system, i.e. $\hat{C}= \hat{H}_S + \hat{p}_t$, it can be shown that it gives the time evolution according to the Schr\"{o}dinger equation, i.e. $\bra{t_2}\hat P\ket{t_1}=e^{-i(t_2-t_1)H_S}$.\\ 

\begin{figure}
\includegraphics[width=0.65\linewidth]{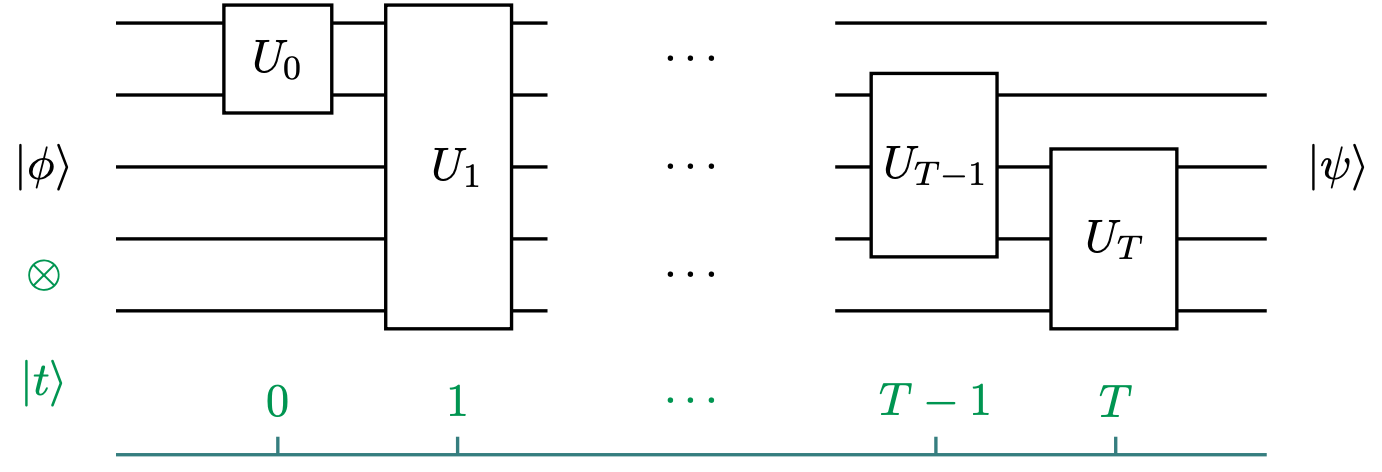} 
\caption{The main idea of a Page-Wootters formulation of a quantum circuit. One considers a quantum clock that keeps track of the number of computational steps that have happened so far. At computational step $t$, the circuit applies the gate $U_t$. The input to the quantum circuit is $\ket{\phi}$ and the output of the circuit is $\ket{\psi}=U_T \cdots U_0 \ket{\phi}$.}
\label{Fig:timeless_circuit}
\end{figure}

The Page-Wootters formalism has been adapted to regular (i.e. causal) quantum circuits, see for example \cite{feynman1985quantum,kitaev2002classical,Breuckmann2014, Caha2018}.  It uses one finite dimensional quantum clock and is described by the constraint equation  
\begin{align}
\hat{C}\ket{\Psi}\rangle =\sum_t \hat{H}_t \ket{\Psi}\rangle =  0, \label{WheelerDeWitt}
\end{align}
where the Hamiltonians 
\begin{align}
\hat{H}_t= -\frac{1}{2}\left( \ketbra{t}{t-1}\otimes U_t  + \ketbra{t-1}{t} \otimes U^{\dagger}_t  -\proj{t-1} -\proj{t} \right) \label{H_t},
\end{align}
can be understood as making the clock tick once and applying some unitary $U_{t}$ to the system. In other words, at time $t$ the circuit applies gate $U_t$. Solutions to Eq.~\eqref{WheelerDeWitt} are history states of this quantum circuit in the form
\begin{align}
\kket{\Psi}= \frac{1}{\sqrt{T+1}}\sum_{t=0}^{T}\ket{t}_C \ox U_t\dots U_0\ket{\phi}_S= \sum_{t=0}^{T}\ket{t}_C \ox \ket{\psi(t)}_S,
\label{Historystate}
\end{align}
with $\ket{\phi}\in \mathcal{H}_S$ being the circuit's input, see Fig.~\ref{Fig:timeless_circuit}. When projecting the clock onto the final time the system is in the state $ \ket{\psi}=U_T\dots U_1\ket{\phi}$, which corresponds to the output of the circuit under consideration. While it is not straightforward to write a physical projector analogous to the continuous case in Eq.~\eqref{Pphys}, we can define a projection operator onto the space of solutions to the constraint equation by
\begin{align}
\hat{P} := \sum_i \ket{\Psi_i} \rangle \langle \bra{\Psi_i}, 
\label{PphysD}
\end{align}
where the $|\Psi_i \rrangle$ are given according to Eq.~\eqref{Historystate} with initial states $|\phi_i\rangle$ taken from an orthonormal basis for $\HH_S$. $\hat{P}$ is the projector onto the space of physical states, which contrarily to the continuous case is now a proper subspace of $\HH_c \ox \HH_S$. Note that similarly to the continuous case we can relate the physical projector to the unitary evolution of the circuit between the respective times by 
\begin{align}
\bra{t_2}\hat P\ket{t_1} =\frac{1}{T+1}U_{t_2}\cdots U_{t_1+1}. \label{Equation:CircuitUfromP}
\end{align}

In what follows we will associate a discrete clock $c_X$ with each agent $X\in \{ A_1 \dots A_N \}$ which gives rise to history states of the form 
\begin{align}
\kket{\Psi}= \sum_{t_{A_1}=0,\dots t_{A_N}=0 }^{T_{A_1}\dots T_{A_N}} \ket{t_{A_1},\dots t_{A_N}}\otimes \ket{\psi(t_{A_1} \dots t_{A_N})}_{S}
=\sum_{t_{A_1}=0,\dots t_{A_N}=0 }^{T_{A_1} \dots T_{A_N}}\ket{t_{A_1}}\dots \ket{t_{A_N}}\ox M_{t_{A_1} \dots t_{A_N}}\ket{\phi},
\label{Historystatemany}
\end{align}
where $\ket{\phi}$ is the initial state of the system. Intuitively, the matrices $M_{t_{A_1},\dots t_{A_N}}$ encode what happens to the system between the initial time and the time when the collection of clocks shows the respective values. By projecting onto a certain clock state $\langle t_X \kket{\Psi}$ we will obtain conditional or perspectival states that correspond to the state agent $X$ assigns to everything other than their own clock at time $t_X$. In the next section we present what we consider reasonable physical assumptions the conditional states and hence the history state have to fulfill.
We will almost exclusively consider the history states $\kket{\Psi}$ as they explicitly represent the perspectives of the agents and the systems and we can directly impose physical requirements on them. The constraint operator $\hat{C}$ can then be implicitly defined afterwards as an operator that annihilates this family of history states. Whether this constraint operator has a simple form, or has desirable properties such as locality, is an interesting question that is nevertheless not pursued in this paper.\\

%In this paper we want to go beyond the circuit model and single clocks by generalizing the discrete Page-Wootters formalism to indefinite causal structure. In that context, we will consider multiple clocks, each of them associated with an agent. 

%%%%%%%%%%%%%%%%%%%%%%%%%%%%%%%%%%%%%%%%%%%%%%%%%%%%%%%

\section{Process matrices within a timeless formalism}
\label{Process matrices within a timeless formalism}

\subsection{The operational setting and postulates}
\label{Section:OurFormalism}

In this section, we develop our framework that allows to model experiments described by pure process matrices within a generalized Page-Wootters approach. \\%Our starting point is the observation of \citep{Castro-Ruiz2020} that history states involving several agents can sometimes give rise to physics involving quantum causal structures.\\

%As in the process matrix formalism~\citep{OCB}, we use the following operational picture: We imagine some agents $A,B,\ldots$ and each of them has their own lab. A basic assumption is that within each lab, the usual rules of quantum theory are still valid. In addition to the usual process matrix approach, the progress of time within the labs is described by quantum clocks with Hilbert spaces $\HH_{c_A}$, $\HH_{c_B}, \ldots$. We refer to the set of all clock variables collectively as $\HH_c$. We will assume that the quantum clocks are discrete, as motivated in Section~\ref{The Page-Wootters formalism}.\\

\begin{figure}[h!]
\begin{subfigure}{0.44\textwidth}
\includegraphics[width=\linewidth]{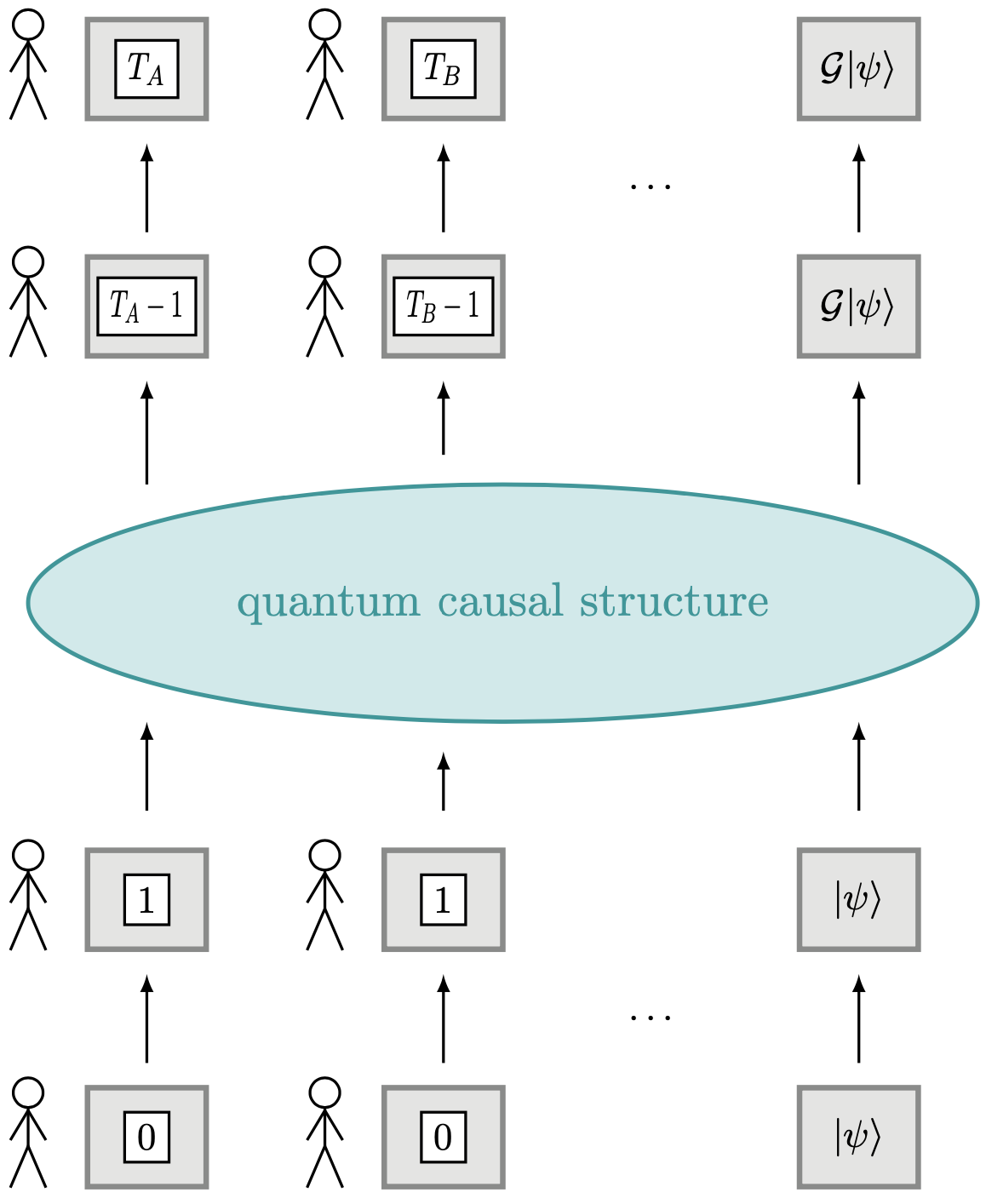} 
\caption{Global perspective}
\label{Figure:Global}
\end{subfigure}
\hfill
\begin{subfigure}{0.44\textwidth}
\includegraphics[width=0.9\linewidth]{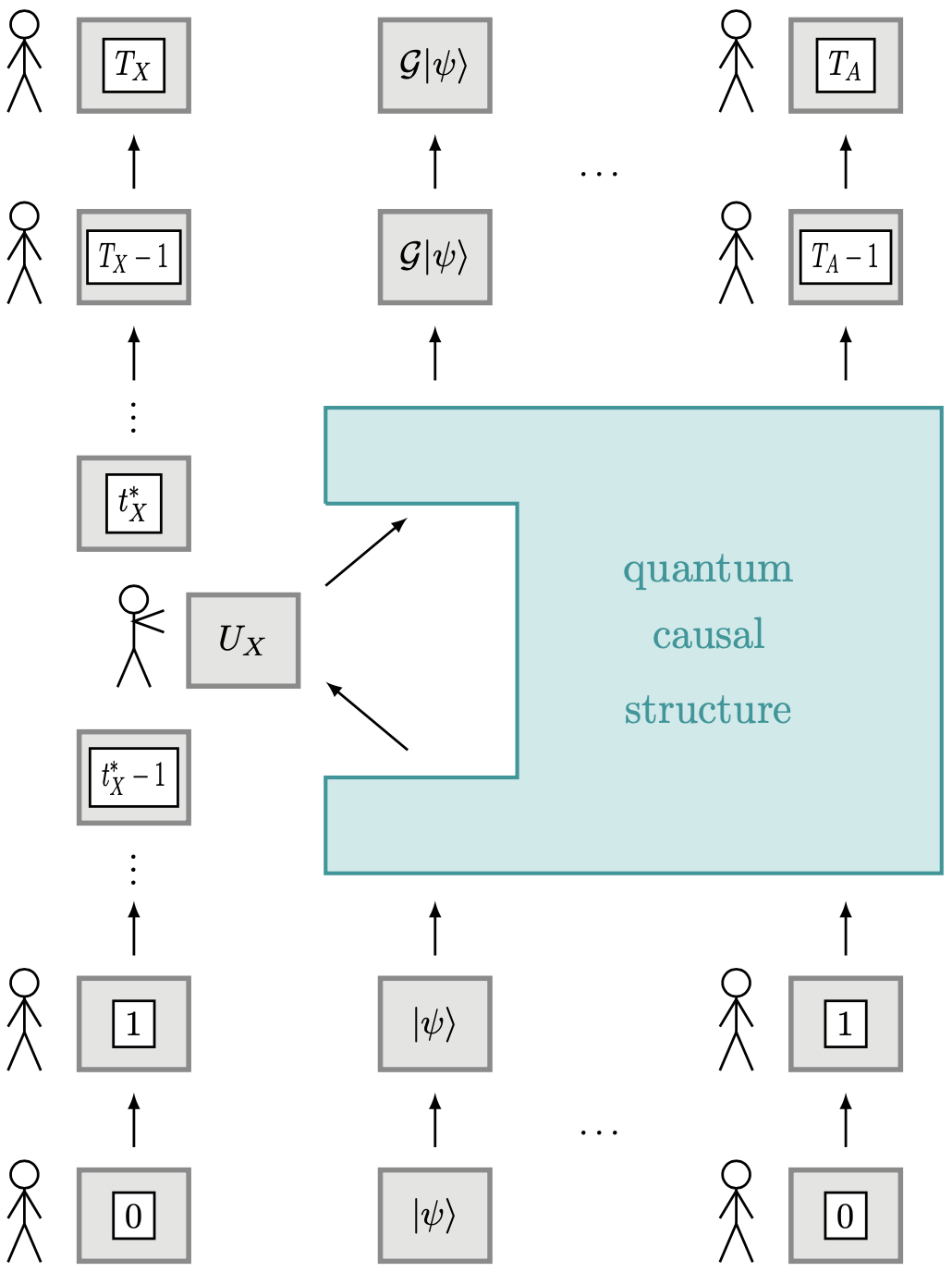} 
\caption{$X$'s perspective}
\label{Figure:Local}
\end{subfigure}
\caption{
The protocol of an experiment involving quantum causal structure from a global (\ref{Figure:Global}) and a local (\ref{Figure:Local}) point of view. At the beginning and end of the experiment, the agents are assumed to be in a definite, space-time causal structure. This is expressed by having their clocks tick in synchronization. However, in between the agents and the main system enter a possibly indefinite causal structure in which the clocks, the main system and the laboratories might get entangled with each other. The ancillary systems for the laboratories are not shown. Inside the laboratories standard quantum theory is valid and therefore each agent $X$, only sees the other agents and the main system as part of a quantum causal structure. At some time $t_X^*-1$, $X$ receives the part of the main system described by $\HH_X$ from the environment. $X$ applies unitary operation $U_X$ to this part of the main system and potential ancillas. Afterwards, $X$ sends that part of the main system back into the environment at time $t_X^*$. The actions of all agents together lead to the process $\mathcal{G}$ being applied to the main system at the end of the protocol.
}
\label{Fig:Framework}
\end{figure}

As in the pure process matrix formalism we will describe scenarios with multiple agents being parts of a well-defined standard global causal past and future. In addition to the usual process matrix approach, the progress of time within the agents' laboratoriess is described by quantum clocks with Hilbert spaces $\HH_{c_{A_1}} \dots \HH_{c_{A_N}}$. We refer to the set of all clock variables collectively as $\HH_c$. The idea of a well defined global, causal past and future common to all agents is formalized by the assumption that at the beginning as well as at the end of the protocol all the clocks experience at least one well-synchronized time step, see Fig.~\ref{Figure:Global}. During the protocol, each agent applies his or her quantum instrument on a part of a system which is common to all agents. As done in~\cite{araujo2017purification} we will assume that each agent has access to an ancillary degree of freedom, denoted by Hilbert spaces $\HH_{A_1'} \dots \HH_{A_N'}$ of unspecified dimension, to implement their quantum instrument. This ancillary system allows to represent the quantum instrument as a unitary within our pure history state approach. The ancilla acts as the environment for a dilation and as memory recording measurement outcomes. We assume that the ancilla systems are initialized to $\ket{0}$ that they have trivial time evolution, except at the moment when the corresponding quantum instrument is applied. We collectively label the ancillas as $\HH_{S'} := \HH_{A_1'} \otimes \dots \otimes \HH_{A_N'} $.\\

In addition to the agents, their ancillas and quantum clocks, we also consider another quantum system that participates in the protocol, described by a Hilbert space $\HH_S$. This quantum system represents the degrees of freedom that play an active role in the protocol, but are not directly associated with the agents or their laboratories. As in the formalism for pure processes, we assume that this quantum system is an input to the causal structure from the global past. We will often call it the \emph{main system} and we denote its initial state by $|\psi\rangle_S$. \\
%when all clocks are at $0$ by $|\psi\rangle_S$. \\

Hence our history states live on $ \HH_c \otimes \HH_S \otimes \HH_{S'}$. We assume that the clocks are initialized to time $0$ at the start of the protocol and show times $T_A,T_B, \dots T_N$ at the end. Then our history states can be expanded in the form
\begin{equation}
|\Psi\rangle \rangle = \sum_{t_{A_1}=0,\dots t_{A_N}=0 }^{T_{A_1}\dots T_{A_N}} \ket{t_{A_1},\dots t_{A_N}}_c\otimes \ket{\psi(t_{A_1} \dots t_{A_N})}_{SS'}.
\label{Equation:HistoryState}
\end{equation}

Now we can formalize our requirements on the timeless state describing the protocol depicted in Fig.~\ref{Fig:Framework}. As mentioned before all clocks and ancillas are initialized to the states $\ket{0}$ and, therefore, we can write: %Furthermore, we assume that the main system is in a state $\ket{\psi}_S$ which is an input to the causal structure from the global past. 

\begin{enumerate}
	\item[\bf S.1] $|\psi (0,0,\dots) \rangle = |\psi \rangle_S |0 \rangle_{S'}$, where $|0 \rangle_{S'} = |0\rangle_{A_1'}\otimes |0\rangle_{A_2'} \otimes \dots \otimes \ket{0}_{A_N'} $ is a fixed ancillary state and $|\psi\rangle_S$ is an arbitrary state of the system.
\end{enumerate}

At the beginning and end of the experiment, physics should be given by a standard space-time causal structure. Hence, at the beginning and in the end, we assume the clocks of the agents are well-synchronized. In particular the clocks perform at least one synchronized step before and after they are part of any exotic causal structure. We further assume that during these initial and final well-synchronized time-steps nothing happens to the main system and formulate this in terms of agent $A$ for the sake of readability. Note that this does not conceptually single out agent $A$ but can equally be written for any of the agents.

\begin{enumerate}
	\item[\bf S.2] $\ket{\psi(0,\dots, t_X, \dots)} \ne 0$ only for $t_X=0$ $\forall X\neq A$ and $ \ket{\psi(T_{A_1}, \dots, t_X, \dots )} \ne 0$ only for $t_X=T_X$ $\forall X\neq A$ and furthermore $\ket{\psi(1,1 \dots 1)} = \ket{\psi(0,0,\dots, 0)}$ and \newline $\ket{\psi(T_{A_1}-1,T_{A_2}-1, \dots, T_{A_N}-1 )} = \ket{\psi(T_{A_1},T_{A_2}, \dots, T_{A_N})}$.  
\end{enumerate}

%Analogous to the process matrices formalism, we assume that during the protocol, each agent obtains a quantum input from the environment. The agent then acts on that input with their quantum instrument and sends it out to the environment again. We model the input from the environment as parts of the main system, i.e. we assume that the input to agent $X$ lives on a subspace $\HH_X \subseteq \HH_S$. $X$'s quantum instrument is represented by a unitary $U_X$ which acts on the received part of the main system and $X$'s ancilla, i.e. $U_X$ acts on $\HH_X \otimes \HH_{X'}$. %The other agents are treated analogously. 
Analogous to the pure process matrices formalism described in Section \ref{Process matrices and causal reference frames}, we model the input from the environment as parts of the main system, i.e. we assume that the input to agent $X$ lives on a subspace $\HH_X \subseteq \HH_S$, and $X$'s quantum instrument is described by a unitary $U_X$ which acts on the received part of the main system and $X$'s ancilla, i.e. $U_X$ acts on $\HH_X \otimes \HH_{X'}$. 
Note that different $\HH_X$ do not need to be different or orthogonal, in fact all of them might even be the full main system Hilbert space $\HH_S$.\\

Next, we discuss the aspect of our discrete clocks formalism that differs the most from the continuous case, and justify the introduction of normalisation operators in order to relate the timeless state $\ket{\Psi \rangle}$ with the local perspective of the agents. As in Ref.~\cite{Castro-Ruiz2020} we condition the history state $\ket{\Psi \rangle}$ on $X$'s clock showing time $t_X$ , i.e. $_{c_X}\braket{t_X}{\Psi \rangle}$, to describe what agent $X$ sees at time $t_X$. In principle, the inner product in the kinematical Hilbert space is not necessarily the same as the inner product for the Hilbert space associated to the perspective of agent  $X$~\cite{Hoehn2019}. Indeed in the usual Page-Wooters formalism with infinite dimensional systems, the physical Hilbert space is not a proper subspace of the kinematical Hilbert space; this necessitates to define a new inner product for the perspectival states. Moreover, even in the finite-dimensional setting that we study here, in scenarios involving clocks with varying relative ticking speeds, one runs into normalization issues if one simply uses the kinematical inner product for the perspectival states. To see this, consider (temporarily switching to continous clocks to illustrate our point) the example of the history state $\ket{\Psi \rangle} =\int \mathrm dt_A \ket{t_A}_{c_A} \otimes \ket{2t_A}_{c_B}$ in which one clock runs twice as fast as the other~\footnote{We will not worry here about our use of non-normalisable states in the infinite dimensional case}. For $A$'s perspective we find $_{c_A}\braket{ t_A }{ \Psi \rangle} = \ket{2t_A}$. However, for $B$'s perspective we find 
\begin{align*}
	_{c_B}\braket{t_B}{\Psi \rangle} = \int \mathrm{d} t_A \ket{t_A} \braket{t_B}{2t_A} = \frac{1}{2} \int \mathrm{d} t'_B \ket{\frac{1}{2}\ t'_B} \braket{t_B}{t'_B} = \frac{1}{2} \ket{1/2\ t_B}, 
\end{align*}
where the prefactor $\frac{1}{2}$ comes from the measure via the change of the integration variable. In this example, the clock ticking rates are constant. However, in general the rates might change dynamically and the corresponding prefactor will depend on time. Hence, different agents will need different renormalizations, which in the infinite dimensional case can be accounted for in the definition of the inner products for the perspectival states. For the finite dimensional case with discrete clocks this motivates the introduction of normalization operators $N_{t_X}^{(X)}$ in order to relate the normalization of the bipartite history state with the normalization of the time-dependent perspectival states. Normalization issues for discrete clocks related to the process of discretization itself are discussed in detail in Appendix \ref{Appendix:Normalization}.\\

%Furthermore, we wish to include physical scenarios involving dynamical clocks with varying clock ticking speeds as they appear in general relativity.
%\textcolor{red}{However, %as shown in Appendix \ref{Appendix:Normalization} one runs into normalisation issues.
%if one considers history states as sums over time instead of integrals in such scenarios, prefactors related to time-dilation need to be taken into account additionally, see Appendix \ref{Appendix:Normalization}. Furthermore, while discretizing clocks, e.g. via time-binning, one quickly runs into normalization issues. A time-binning would typically map close states $\ket{t}$ and $\ket{t+\delta t}$ to the same discrete time state, changing the normalization. }  
%We compensate for such artifacts of discretization by introducing normalization operators $N_{t_X}^{(X)}$.\\
We assume that the state agent $X$ sees at time $t_X$ is 
\begin{equation}
|\psi_X(t_X) \rangle = N^{(X)}_{t_X} \braket{t_X}{\Psi \rangle} = \langle t_X|_{c_X} \otimes N^{(X)}_{t_X}|\Psi \rrangle,
\label{perspectival}
\end{equation}
where $ N^{(X)}_{t_X} \in \LL(\HH_{c_{\backslash X}} \otimes \HH_S \otimes \HH_{S'})$ is the normalization operator that relates the perspective-neutral description to the perspective of agent $X$ at time $t_X$. Here, $\HH_{c_{\backslash X}}$ is the Hilbert space formed by all clocks except the clock of agent $X$.  \\

A priori, the normalization operators make this approach extremely general. In principle, they could give us any state $\ket{\psi_X(t_X)}$ that we want. Therefore it is important that we impose some extra conditions. First of all, as the normalization operators generalize normalization constants, they should be linear, positive and invertible. Moreover, we wish that all the relevant physics concerning the initial system state $\ket{\psi}_S$ and the agents' operations is encoded in the history state, not the normalization operators. The normalization operator should just correct the normalization depending on the clocks. Therefore, we demand that the operators $N_{t_X}^{(X)}$ are independent of the initial system state $\ket{\psi}$ and the choice of quantum instruments by the agents.

\begin{enumerate}
\item[\bf N.1] $N^{(X)}_{t_X} $ is an invertible, linear, positive operator. It is independent of the input state $\ket{\psi}_S$ and the local operations $U_{A_1}\dots U_{A_N}$.
\end{enumerate}

Without the latter restriction, one could use $N_{t_X}^{(X)}$ to introduce copies of the initial state $\ket{\psi}_S$ or apply copies of the agents' instruments to violate the no-cloning principle.
We further assume that the normalization operator does not perturb how one agent sees the clocks of the other agents: 
\begin{enumerate}
\item[\bf N.2] The normalization operator has the form
\begin{equation}
N_{t_X}^{(X)} = \sum_{t_{A_1}, \dots, \widehat{t_X}, \dots t_{A_N}} \proj{t_{A_1}, \dots \widehat{t}_X,\dots t_{A_N}} \otimes n_{t_{A_1}, \dots \widehat{t}_X,\dots t_{A_N}}^{(X)} \otimes \id_{S'} 
\end{equation}
where the sum is taken over all clocks except the clock of agent $X$, which is omitted, as indicated by $\widehat{t_X}$. The operator $n_{t_{A_1}, \dots \widehat{t}_X,\dots t_{A_N}}^{(X)}$ is a linear, invertible and positive operator acting on $\HH_S$ (but not on the ancillas $\HH_{S'}$).
\end{enumerate}
This assumption is motivated by the requirement that the history state should represent the relevant physics and relations of the clocks. As their name implies, the normalization operators should adjust the normalization, but not introduce new clock physics. 

Our previous requirement of well-synchronized clocks at the beginning and end of the experiment additionally implies that the respective normalization operators should just be identity operators. 
\begin{enumerate}
\item [\bf N.3] $N_{1}^{(X)} = N_{0}^{(X)} = \mathbb 1$ as well as $N_{T_X-1}^{(X)} = N_{T}^{(X)} = \mathbb 1$ $\forall X $.
\end{enumerate}

Finally, we have to explain how the perspectival states $N_{t_X}^{(X)}\braket{t_X}{\Psi \rangle}$ are related to each other. We will assume that each agent $X$ sees a unitary time evolution as dictated by quantum theory in a pure state framework.
This means we assume that for all $t_X, t'_X$ there exists a unitary operator $\UU_X(t_X,t'_X)$ such that
\begin{equation}
\ket{\psi_X(t_X)} = \UU_X(t_X,t'_X) |\psi_X(t_X') \rangle.
\end{equation}
Furthermore, just like in usual quantum theory, $\UU_X(t,t')$ should not depend on the initial system state $\ket{\psi}_S$. 
\begin{enumerate}
\item[\bf U.1] $\UU_X(t, t')$ is a unitary operator, independent of the initial state $|\psi \rangle_S$.
\end{enumerate}
Moreover, time-evolution from $t''$ to $t'$ to $t$ is the same as time-evolution from $t''$ to $t$.
\begin{enumerate}
\item[\bf U.2] $\UU_X(t, t') \UU_X(t', t'' ) = \UU_X(t,t'')$, $\forall t ,t', t''$.
\end{enumerate}
Next we discuss the crucial assumption that connects our framework to the formalism of pure processes. In the process matrix framework, one assumes that during the protocol each agent eventually receives a quantum system from the environment. We will assume that each agent is promised that they will receive this quantum system at a specific time $t_X^*-1$. As explained before, we model that quantum system to be a part of the main system, described by subspaces $\HH_{A_1} \dots \HH_{A_N}$  of the main system space $\HH_S$. Each agent $X$ acts with their local operation $U_X$ on that system and their ancilla (i.e. $U_X \in \mathcal L(\HH_X \otimes \HH_{X'})$) and then sends out the system at $t_X^*$.  In particular, this is the only time the agents use their quantum instrument. While the agents enforce evolution via their instrument, the remaining degrees of freedom should evolve in an uncorrelated way. This leads us to our final requirement, which is analogous within our framework to the existence of a causal frame decomposition Eq.~\eqref{Equation:GfromCausalRefFrames} of process matrices.
\begin{enumerate}
\item[\bf U.3] $X$'s quantum instrument is used at the so called \emph{time of action} $t_X^*$, i.e.
\begin{equation}
	\UU_X(t_X^*,t_X^* -1 ) = U_X \otimes \mathrm{Rest}^{(X)}.
\end{equation}
Furthermore at other times $t \neq t_X^*$ the evolution operator $\UU_X(t, t-1)$ is independent of $U_X$ and only acts as the identity on the ancilla of $X$, i.e. on $\mathcal{H}_{X'}$. 
\end{enumerate}

\noindent
Our assumptions introduce a transformation that maps the initial state $\ket{\psi(0,\dots,0)}$ to the final state $\ket{\psi(T_{A_1},  \dots, T_{A_N})}$. This transformation depends on the agents' actions $U_X$ and is visualized in Fig.~\ref{Fig:Framework}. Our last assumption is that this transformation can be extended to a full process~\cite{araujo2017purification} i.e. quantum superchannel~\cite{TulioSuperchannels}. This means that it must be possible to interpret the quantum causal structure as a process, even if we describe the agents' operations as channels instead of (purified) unitaries.
\\

We make the assumption that our Postulates \textbf{S.2}, \textbf{N.1}, \textbf{N.2}, \textbf{N.3}, \textbf{U.1}, \textbf{U.2} and \textbf{U.3} continue to be satisfied if the ancillary systems are initialized to states other than $\ket{0}_{A'_j}$, and that we can continue to use the same normalization operators $N_{t_X}^{(X)}$ and perspectival time evolutions $\mathcal U_{X}(t_X+1, t_X)$ as for the initialization $\ket{0,\dots,0}_{S'}$. This is no substantial conceptual restriction, because none of these postulates explicitly refers to any particular initial ancillary system state. Postulate \textbf{S.1} just defines the particular choice of initialization for the protocol. \\

This concludes the description of the operational setting and of our assumptions. We will subsequently investigate the mathematical and physical implications of our setting and postulates.

\subsection{History states lead to pure processes}
\label{Section:HistoryPure}
%In this subsection we will investigate the mathematical and physical implications of our setting and postulates. We will start by showing that the causal structure arising in our framework is indeed described by pure process matrices. In that context we will explain that our formalism refines the causal reference frame formalism by explicitly modeling the quantum clocks. In particular, as we will see, our history state approach adds further constraints on the causal reference frames. At last, we will formally introduce a physical projector onto the set of history states and a corresponding constraint operator.\\

First, we show that the evolution of the main system and the ancillas must be given by a pure process. For that purpose we have to analyze the relation between the initial and the final state, in particular with respect to the operations of the agents. This can be done by taking the perspective of an agent, for example $A_1$, and applying our unitary time evolution postulates:
\begin{align}
&\ket{T_{A_2}\dots T_{A_N}}_{c_{\backslash A_1}} \ox \ket{\psi(T_{A_1},T_{A_2},\dots T_{A_N})} = \nonumber \\
&\qquad \qquad \left(\UU_{A_1}(T_{A_1},t_{A_1}^*) (U_{A_1} \otimes \mathrm{Rest}^{(A_1)}) \UU_{A_1}(t_{A_1}^*-1,0)\right) \big( \ket{0, \dots 0}_{c_{\backslash A_1}} \ox \ket{\psi(0,0,\dots 0)} \big) \label{Equation:Gappears} %\\
%& = \ket{T_B,\dots T_N}\otimes \mathcal{G}(U_A,U_B \dots) \ket{\psi(0,0,\dots )}, \nonumber
\end{align} 

We can define a map $\mathcal G$ that describes how the final system and ancilla state is related to the initial state:

\begin{align}
&\ket{\psi(T_{A_1}\dots T_{A_N})} =: \mathcal G(U_{A_1}\dots U_{A_N}) \ket{\psi(0,0,\dots 0)}. \label{Equation:DefG}
\end{align}

Eq.~\eqref{Equation:Gappears} shows that $\mathcal G(U_{A_1}\dots U_{A_N}) $ is a unitary that maps the initial system and ancilla state to the final state and that it is multilinear in the local operations. Furthermore, this decomposition shows that the only change in the state of the ancilla of $A_j$ is caused by $A_j$'s local operation. We assumed that this map can be extended to a full process. We conclude that $\mathcal G$ is a pure process as defined in \citep{araujo2017purification,TulioSuperchannels} with the difference that Eq.~\eqref{Equation:Gappears} represents a refined causal reference frame decomposition that explicitly includes the quantum clocks, compare to Eq.~\eqref{Equation:GfromCausalRefFrames}.

The fact that we obtain pure processes has important consequences: According to \cite{BarrettOreshkovCausalModels,TulioSuperchannels}, in the bipartite case our setting implies that no violation of device-independent causal inequalities can occur: The bipartite pure process can only be causally ordered or quantum-controlled causal order. \\
%%%%%%%%%%%%%

\subsection{Additional restrictions for the local perspectives of the agents}
\label{Section:AffineLinearM}

Let us further investigate the relation between our framework and the original causal reference frame framework of~\cite{guerin2018agent}, in particular the relation between Eqs. \eqref{Equation:GfromCausalRefFrames} and \eqref{Equation:Gappears}, in further detail. %As explained in Section \ref{Process matrices and causal reference frames} the causal reference frame allows to describe the perspective of an agent inside a causal structure. Also our framework describes the perspective of an agent. Therefore it is important the we compare these two frameworks and investigate their compatibility. \\%As we explicitly model the clocks by using history states, we can $\ket{\Psi_X(t_X)} = N_{t_X}^{(X)} {_{c_X}\braket{t_X}{\Psi \rangle}}$. \\
%First, let us state the similarities of the two formalisms: 
Both frameworks work with purifications, in particular the actions of the agents are described by purified unitaries $U_{A_1}\dots U_{A_N}$ and the relevant process matrices turn out to be the pure processes. The crucial objects of the causal reference frame framework are the unitaries that describe the past before and the future after an agent's action. More specifically, from the point of view of agent $X$, the evolution from the beginning of the protocol up to the time of $X$'s action is described by the unitary $\Pi_X$. In our framework, this unitary corresponds to $\UU_X(t^*_X-1,0)$. The evolution directly after $X$'s action up to the end of the protocol is described by $\Phi_X$, which in our framework corresponds to $\UU_X(T_X,t^*_X)$. 

The crucial difference between our framework and that of causal reference frames is that we explicitly model the quantum clocks and explain how the agents' perspectives arise from a perspective neutral history state. This gives us a refined description of the agents' perspectives because we explicitly model individual time steps $t_X \rightarrow t_X+1$ in between the beginning of the experiment, the time of action and the end of the protocol. In Ref.~\cite{guerin2018agent} the causal past and future unitaries $\Phi_X$ and $\Pi_X$ are allowed to be arbitrary as long as they combine to the pure process $\mathcal G$ via Eq.~\eqref{Equation:GfromCausalRefFrames}. However, in our setting the history state induces further compatibility constraints on the perspectives of the agents. 

We will now present one such constraint that is particularly restrictive: Affine-linearity in the operations of the other agents. Consider a history state as in Eq.~\eqref{Equation:HistoryState}. We can write
\begin{align}
	\ket{\psi(t_{A_1}\dots t_{A_N})} = M_{t_{A_1}\dots t_{A_N}} \ket{\psi(0,0,\dots 0)} \label{Eq:MrelatesStates}, 
\end{align}
with 
\begin{align}
	M_{t_{A_1} \dots t_{A_N}}  =  {}_{c_{\backslash A_1}}\bra{t_{A_2},\dots t_{A_N}} (N_{t_{A_1}}^{(A_1)})^{-1}\UU_{A_1}(t_{A_1},0)\ket{0,\dots 0}_{c_{\backslash A_1}} \label{Equation:MfromU}.
\end{align}
Similar equations hold for the other agents. From our assumptions, we can see that $M_{t_{A_1}\dots t_{A_N}}$ is constant in $U_{A_1}$ for $t_{A_1} < t_{A_1}^*$ and linear in $U_{A_1}$ for $t_{A_1} \ge t_{A_1}^*$, because the same is true for $\UU_{A_1}(t_{A_1},0)$.
We can relate the time evolutions of two different agents (here $A_1$ and $A_2$) via
\begin{align}
	&\UU_{A_2}(t_{A_2},0) \ket{0,0,\dots 0}_{c_{\backslash A_2}}= \sum_{t_{A_1} ,t_{A_3},\dots t_ {A_N}}N_{t_{A_2}}^{(A_2)} \ket{t_{A_1},t_{A_3},\dots t_{A_N}} _{c_{\backslash A_2}} M_{t_{A_1},\dots t_{A_N}} 
\label{Eq:relating_perspectives}
\end{align}
The dependence of $ M_{t_{A_1},\dots t_{A_N}} $ on $U_{A_1}$ shows that $\UU_{A_2}(t_{A_2},0)\ket{0,0,\dots}$ is a sum of functions linear in $U_{A_1}$ or constant in $U_{A_1}$, i.e. $\UU_{A_2}(t_{A_2},0)\ket{0,0,\dots}$ is affine-linear in $U_{A_1}$. The same argument can be made for all other agents. Hence we get that any time evolution $\UU_X(t_X,0)$ as seen by agent $X$ (with all clocks initialized to time $0$) has to be an affine linear function of the operations of all other agents. This affine linearity is a severe restriction and a potential obstacle for implementing some non-causal processes in this framework. In Section \ref{sec:Lugano process} we will apply this insight to an example involving an exotic tripartite process~\cite{araujo2017purification,baumeler2016space} to see that a causal reference frame decomposition for this process in Ref.~\cite{guerin2018agent} is incompatible with our framework. \\%because it involves a gate that is not affine-linear in the operations of the other agents.\\

%%%%%%%%%%%%%
\subsection{Discrete constraint operators and physical projectors}
\label{Section:ConstraintOperator}

Finally, we will briefly discuss constraint operators and physical projectors in our framework since they are among the main objects of interest in the Page-Wootters formalism presented in Section \ref{The Page-Wootters formalism}.
%So far, we focused on the history states. However, as explained in Section \ref{The Page-Wootters formalism}, in the original Page-Wootters formalism as well as its adaption to circuits one often considers constraint operators and physical projectors. As mentioned earlier, it is possible to introduce such operators in our setting too, as we will now explain in detail. 
By construction, our history states form a subspace $\HH_H \subset \HH_c \otimes \HH_S \otimes \HH_{S'}$ and by linearity, $\alpha \ket{\Psi\rangle} + \beta \ket{\Psi' \rangle}$ is the history state associated with the input state $\alpha\ket{\psi}_S + \beta \ket{\psi'}_S$, as one can see e.g. from Eq.~\eqref{Eq:MrelatesStates}. Therefore, we can define a constraint operator $\hat{C}$ as $\hat{C} = \mathbb 1 - \hat{P}_{H}$ where $\hat{P}_H$ is the orthogonal projector onto $\HH_H$. Then the kernel of $\hat{C}$ is given by $\HH_H$.

We note that in general, $\hat{P}_H$ in our framework cannot be written analogous to the case of a standard quantum circuit with one clock, see Eq.~\eqref{PphysD}. More specifically, for an orthonormal basis $\ket{\psi_j}_S$, the corresponding history states 
\begin{align}
	\ket{\Psi_j \rangle} &= \sum_{t_{A_1}\dots t_{A_N}} \ket{t_{A_1}\dots t_{A_N}} \ox \ket{\psi_j(t_{A_1}\dots t_{A_N})}_S = \sum_{t_{A_1} =0}^{T_{A_1}} \ket{t_{A_1}} (N^{(A_1)}_{t_{A_1}})^{-1} \ket{\psi_{A_1,j}(t_{A_1})} \nonumber \\	
	&= \sum_{t_{A_1} =0}^{T_{A_1}} \ket{t_{A_1}} (N^{(A_1)}_{t_{A_1}})^{-1}\UU_{A_1}(t_{A_1},0) \ket{0,0,\dots} \ox \ket{\psi_j}_S  \nonumber
\end{align}
may fail to be orthogonal due to the normalization operators:
\begin{align}
	\braket{\langle \Psi_k}{\Psi_j \rangle} = \sum_{t_{A_1} =0}^{T_{A_1}}\bra{\psi_k}_S  \ox \bra{0,0,\dots 0} \UU_{A_1}(t_{A_1},0)^\dagger[(N^{(A_1)}_{t_{A_1}})^{-1}]^\dagger  (N^{(A_1)}_{t_{A_1}})^{-1} \UU_{A_1}(t_{A_1},0) \ket{0,0,\dots 0} \ox\ket{\psi_j}_S  . \nonumber
\end{align}
In that sense, the map from initial states to history states is not necessarily unitary, in contrast to the unitary evolution of the main system and ancilla state, see Eq.~\eqref{Equation:DefG}. 

We can, however, write $\hat{P}_H$ in a form more reminiscent of the original Page-Wootters framework, compare Eq. \eqref{Pphys}, as
\begin{equation}
\hat{P}_H = \frac{1}{T} \sum_{k=0}^{T-1} \exp\left( -2 \pi i \hat{C} \frac{k}{T}\right),
\label{Eq:P_H}
\end{equation}
where $T$ is an integer (we could take $T= T_{A_1}$).
This can be seen by noting that $\hat{C}$ is a hermitian matrix with only eigenvalues $0$ or $1$. If $|\phi_0\rangle$ is an eigenvector of $\hat{C}$ with $\hat{C} |\phi_0\rangle = 0$, we have $\hat{P}_H |\phi_0\rangle = \ket{\phi_0}$,  while if $\hat{C} |\phi_1 \rangle = \ket{\phi_1}$ we have $\hat{P}_H |\phi_1 \rangle = \frac{1}{T} \sum_{k=0}^{T-1} e^{ -2 \pi i \frac{k}{T}} \ket{\phi_1}= 0$, showing $\hat{P}_H = \mathbb 1 - \hat{C}$. As discussed in Appendix \ref{App:P} it is not clear whether $\hat{P}_H$ in Eq.\eqref{Eq:P_H} can be linked to the perspectival unitaries $\UU_X(t'_X,t_X)$ similar to Eq.~\eqref{Equation:CircuitUfromP}.

%$P_H$ defines a projection operator into the set of history states and has a form reminiscent of the original Page-Wootters framework. However, 

%\subsection{Relation to causal reference frames}
%\label{Section:RelationToCausalReferenceFrames}

\section{Causal and non-causal Page-Wootters circuits}
\label{Non-causal Page-Wootters circuits}

In this section we now apply our framework to give several examples of physical scenarios that go beyond the standard setting of circuits with well-synchronized clocks. First, in Section~\ref{"Twin-paradox" circuit}, we consider a setup inspired by the famous twin paradox in which the clocks of two agents are still in a well-defined relation to each other, but tick at different rates. Afterwards, in Section~\ref{sec:switch}, we describe a scenario for the bipartite quantum switch as a prototypical example of a known class of non-causal processes. There, a control quantum system determines the tick rates of the agents' clocks and more importantly the order of the agents' operations. In Section~\ref{Section:ControlledCombs}, we go beyond the example of the bipartite switch and show that arbitrary coherently controlled causal orders can be realized in our framework. Finally, in Section~\ref{sec:Lugano process}, we consider an interesting pure, non-causal process that is further known to violate causal inequalities~\cite{baumeler2016space, baumeler2014maximal, araujo2017purification}. We will argue that this process cannot be implemented as a superposition of classical histories and that the causal reference frame decomposition from Ref.~\cite{guerin2018agent} cannot be adapted to our setting.

\subsection{A history state for a scenario with varying clock ticking rates}
\label{"Twin-paradox" circuit}

%In this section, we construct a first example of an interesting process that fits into our setting but is not a standard circuit. Specifically we consider a scenario that features varying clock speeds, inspired by the famous twin paradox. \\
Our first example of an interesting process that fits into our setting but is not a standard circuit is inspired by the famous twin paradox. Specifically we consider a scenario that features varying clock speeds of two agents $A$ and $B$ where during the protocol the clock of one ticks slower than the clock of the other, reminiscent of the one twin that leaves earth traveling at relativistic speed and returns to find his or her sibling older than they are themselves.\\

\begin{figure}[h!]
\includegraphics[width=0.5\linewidth]{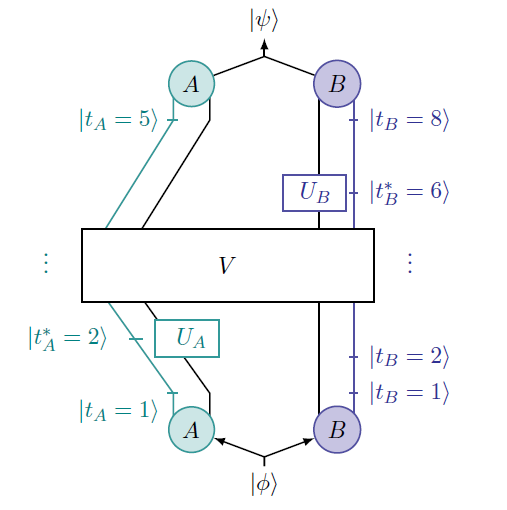} 
\caption{Example of a setting involving a clock with changing ticking rate. The two agents $A$ and $B$ each receive a part of the input system and experience one synchronized time step. After that the clock of $A$ starts ticking slower and $A$ applies their unitary $U_A$ to their subsystem of state $\ket{\phi}$. This is followed by some unitary evolution $V$ of the full system, which is independent of the two agents. Then $B$ applies their unitary $U_B$ to his or her subsystem and finally, at the end of the protocol, the two clocks tick in synchronization once more.}
\label{Fig:Twin-paradox_circuit}
\end{figure}

Here the two agents act on subsystems $S_A$ and $S_B$ of the input quantum state $\ket{\phi} \in \HH_S$ with unitary operations $U_A$ and $U_B$ respectively. The causal order in this example is well defined and we consider the case where $A$ acts before $B$. Moreover, between $A$'s and $B$'s time of action some global evolution $V$ of the system happens, which is independent of the two agents. In the beginning and at the end the agents' clocks tick at the same speed, but in between the clock of $A$ ticks more slowly than that of $B$. The scenario is depicted in Figure~\ref{Fig:Twin-paradox_circuit} and captured by the following history state $\kket{\Psi} \in \HH_{c_A} \otimes \HH_{c_B} \otimes \HH_{S_A} \otimes \HH_{S_B}$.
%We consider agents $A$ and $B$, with systems $S_A$ and $S_B$ that they will act on with their operations $U_A$ and $U_B$. The systems are intially in the joint state $\ket{\phi}$. In the beginning and end the agents' clocks tick at the same speed, but in between the clock of $B$ ticks faster than that of $A$, see Fig.~\ref{Fig:Twin-paradox_circuit}. First $A$ acts with $U_A$, then a global evolution $V$ happens, afterwards $B$ acts with $U_B$. Such a situation is captured by the physical state 
\begin{align}
\kket{\Psi}=&\ket{0_A,0_B}_c\ox \ket{\phi}
+\ket{1_A,1_B}_c \ox \ket{\phi} +\ket{2_A,2_B}_c\ox (U_A \otimes \mathds{1}) \ket{\phi} +\ket{2_A,3_B}_c\ox (U_A \otimes \mathds{1}) \ket{\phi} +\ket{3_A,4_B}_c\ox V(U_A \otimes \mathds{1} )\ket{\phi}   \nonumber \\
& +\ket{3_A,5_B}_c\ox V(U_A \otimes \mathds{1})\ket{\phi} +\ket{4_A,6_B}_c\ox ( \mathds{1} \otimes U_B ) V(U_A \otimes \mathds{1})\ket{\phi} +\ket{4_A,7_B}_c\ox \mathcal{G}(U_A,U_B)\ket{\phi} \nonumber\\
&+\ket{5_A,8_B}_c\ox \mathcal{G}(U_A,U_B)\ket{\phi} +\ket{6_A,9_B}_c\ox \mathcal{G}(U_A,U_B)\ket{\phi}
\label{historyCircuit}
\end{align}
where $\mathcal{G}(U_A,U_B)=( \mathds{1} \ox U_B )  V(U_A \ox\mathds{1})$. The perspectival states for the two agents including the non-trivial normalization operators are

\begin{align}
&\ket{\psi_A(0)}=\ket{0_B}_{c_B}\ox \ket{\phi}, \qquad \ket{\psi_A(1)}=\ket{1_B}_{c_B}\ox \ket{\phi}, &&\ket{\psi_B(0)}=\ket{0_A}_{c_A}\ox \ket{\phi}, \qquad \ket{\psi_B(1)}=\ket{1_A}_{c_A}\ox \ket{\phi}, \nonumber \\
&\ket{\psi_A(2)}=\frac{1}{\sqrt{2}}(\ket{2_B}+\ket{3_B})_{c_B}\ox (U_A \ox\mathds{1})\ket{\phi},
&&\ket{\psi_B(2)}=\ket{2_A}_{c_A}\ox  (U_A \ox\mathds{1} )\ket{\phi} \nonumber \\
&\qquad \text{with }N_2^{(A)}=\frac{1}{\sqrt{2}}\mathds{1}_S,
&&  \quad\qquad =\ket{\psi_B(3)}, \nonumber \\
&\ket{\psi_A(3)}=\frac{1}{\sqrt{2}}(\ket{4_B}+\ket{5_B})_{c_B}\ox V(U_A \ox \mathds{1})\ket{\phi}
&&\ket{\psi_B(4)}=\ket{3_A}_{c_A}\ox  V(U_A\ox\mathds{1})\ket{\phi}  \label{Equation:TwinPerspectives} \\
&\qquad \text{with }N_3^{(A)}=\frac{1}{\sqrt{2}}\mathds{1}_S,
&& \quad \qquad=\ket{\psi_B(5)}, \nonumber \\
&\ket{\psi_A(4)}=\frac{1}{\sqrt{2}}(\ket{6_B}+\ket{7_B})_{c_B}\ox ( \mathds{1} \ox U_B) V(U_A\ox \mathds{1})\ket{\phi}
&&\ket{\psi_B(6)}=\ket{4_A}_{c_A}\ox  ( \mathds{1} \ox U_B )  V(U_A \ox \mathds{1})\ket{\phi} \nonumber \\
&\qquad \text{with }N_4^{(A)}=\frac{1}{\sqrt{2}}\mathds{1}_S,
&& \qquad \quad =\ket{\psi_B(7)}, \nonumber \\
&\ket{\psi_A(5)}=\ket{8_B}_{c_B}\ox \mathcal{G}(U_A,U_B)\ket{\phi},
&&\ket{\psi_B(8)}=\ket{5_A}_{c_A}\ox \mathcal{G}(U_A,U_B)\ket{\phi}, \nonumber \\
&\ket{\psi_A(6)}=\ket{9_B}_{c_B}\ox \mathcal{G}(U_A,U_B)\ket{\phi}, &&\ket{\psi_B(9)}=\ket{6_A}_{c_A}\ox \mathcal{G}(U_A,U_B)\ket{\phi}. \nonumber
\end{align}

Note that the normalization operators are non trivial for precisely those times where the clock of $A$ ticks slower than the clock of $B$. The states in Eqs.\eqref{Equation:TwinPerspectives} can be reproduced by the following unitary evolutions with respect to the two agents 
\begin{align}
& &&\mathcal{U}_B(1,0)=T_{c_A}\ox \mathds{1}_S, &&  \nonumber\\
& \mathcal{U}_A(1,0)=T_{c_B}\ox \mathds{1_S},
&&\mathcal{U}_B(2,1)= T_{c_A}\ox (U_A \ox \mathds{1} )_S,\nonumber\\
& \mathcal{U}_A(2,1)=(T'_{2})_{c_B}\ox (U_A \ox \mathds{1})_S,
&&\mathcal{U}_B(3,2)= \mathds{1}, \nonumber \\
&\mathcal{U}_A(3,2)= (T^2)_{c_B}\ox V_S,
&&\mathcal{U}_B(4,3)= T _{c_A}\ox V _S\nonumber, \\
& \mathcal{U}_A(4,3)= (T^2)_{c_B}\ox ( \mathds{1} \ox U_B)_S ,
&&\mathcal{U}_B(5,4)= \mathds{1}, \label{Equation:UnitariesTwin}\\
& \mathcal{U}_A(5,4)= (T'_{6})_{c_B}\ox \mathds{1}_S,
&&\mathcal{U}_B(6,5)= T_{c_A}\ox (\mathds{1} \ox U_B )_S, \nonumber \\ 
& \mathcal{U}_A(6,5)=T _{c_B}\ox \mathds{1}_S, 
&& \mathcal{U}_B(7,6)=\mathds{1} \nonumber \\ & &&  \mathcal{U}_B(8,7)=T _{c_A}\ox \mathds{1}_S=\mathcal{U}_B(9,8), \nonumber
\end{align}
where $T$ is the unitary that makes the clock of the other agent tick, i.e. $T:\ket{t}\mapsto \ket{t+1}$. Similarly, $T'_{i}$ is any unitary that acts as $\ket{i-1}\mapsto 1/\sqrt{2}(\ket{i}+\ket{i+1}), \ 1/\sqrt{2}(\ket{i}+\ket{i+1})\mapsto \ket{i+2}$.
%a unitary that makes the other agent's clock tick, except for the subspace spanned by $\ket{i-1}, \ket{i}, \ket{i+1},\ket{i+2}$ where it acts as follows: $\ket{i-1}\mapsto 1/\sqrt{2}(\ket{i}+\ket{i+1}), \ 1/\sqrt{2}(\ket{i}+\ket{i+1})\mapsto \ket{i+2}$
We see that our axioms are fulfilled and the times of action are $t_A^{\star}=2$ and $t_B^{\star}=6$ respectively. 
As one can see in Eqs.~\eqref{Equation:UnitariesTwin}, from $A$'s perspective $B$'s clock seems to tick at double the rate in the middle of the process, while from the point of view of $B$, $A$'s clock seems partially frozen in time.
%, which mimics the famous twin paradox with $B$ being the twin for whom more time has elapsed when $A$ returns at the end of the protocol, see Fig.~\ref{Fig:Twin-paradox_circuit}.

%=====================================%
\subsection{The quantum switch} 
\label{sec:switch}

\begin{figure}[h!]
\vspace{-1em}
\includegraphics[width=0.75\linewidth]{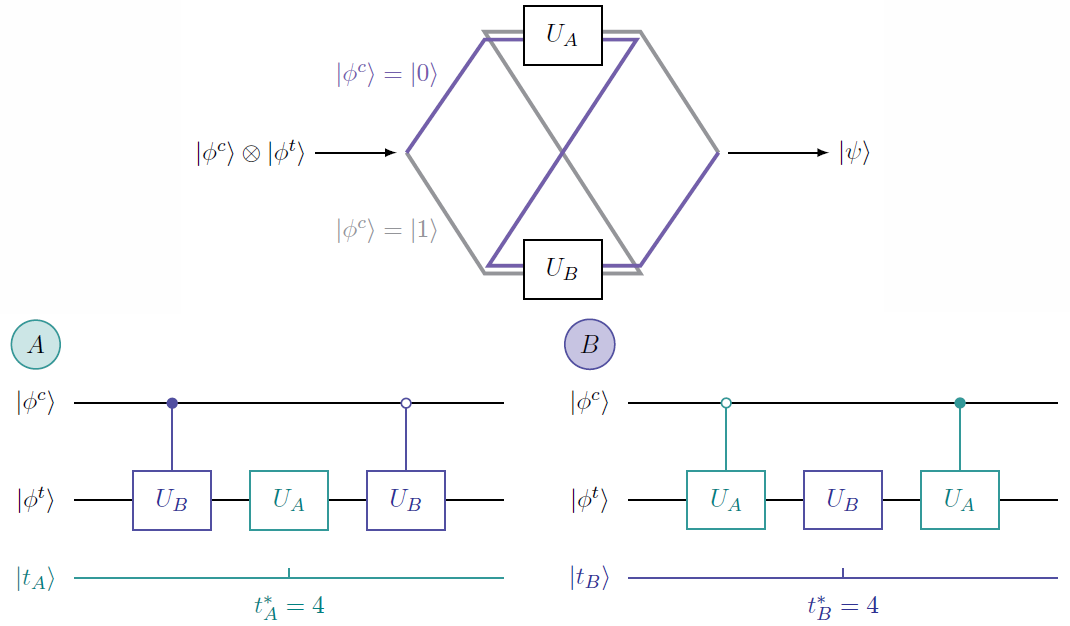} 
\caption{The bipartite quantum: Depending on the value of a control qubit the two unitaries $U_A$, $U_B$ are applied to the target system in different order (top). According to the perspectives of the two agents, $A$ or $B$ apply their own unitary to the target system at time $t^*_A$ or $t^*_B$ respectively, while the other agent's unitary is applied either before or after that depending on the value of the control system (bottom). The perspectival circuits equal the causal reference frames for the quantum switch given in Ref.~\cite{guerin2018agent}.
}
\label{Fig:QSwitch}
\end{figure}

Our second example describes the probably best known non-causal process, namely the bipartite quantum switch~\cite{chiribella2013quantum}. A schematic picture as well as the two perspectival circuits analogous to the causal reference frame decomposition given in Ref.~\cite{guerin2018agent} are shown in Fig.~\ref{Fig:QSwitch}.
%The best known examples of non-causal processes are those where the causal order is coherently controlled, the simplest version of which is the quantum switch~\cite{chiribella2013quantum}, see Figure \ref{Fig:QSwitch}. 
The bipartite quantum switch can be modeled by a history state complying with our axioms starting with an initial state $\ket{\phi}_S \in  \HH_{S_c} \otimes \HH_{S_t}$ consisting of a control ancilla and a target system. Both agents are acting on the target system Hilbert space; $\HH_A = \HH_B = \HH_{S_t}$.

A history state of the quantum switch is given by
\begin{align}
|\Psi \rrangle&=\ket{0_A,0_B}_c \ox \ket{\phi}
+\ket{1_A,1_B}_c \ox \ket{\phi} \nonumber +\ket{2_A,2_B}_c \ox \ket{\phi} +\ket{3_A,2_B}_c \ox (\proj{0}\otimes \mathds{1})\ket{\phi} +\ket{2_A,3_B}_c \ox (\proj{1}\otimes \mathds{1})\ket{\phi}\nonumber \nonumber \\
&
+\ket{4_A,3_B}_c \ox (\proj{0}\ox U_A) \ket{\phi} + \ket{3_A,4_B}_c(\proj{1} \ox U_B) \ket{\phi} +\ket{5_A,4_B}_c \ox (\proj{0}\ox U_B U_A)\ket{\phi} \label{historySWITCH} \\
& + \ket{4_A,5_B}_c \ox(\proj{1} \ox U_A U_B) \ket{\phi} +\ket{5_A,5_B}_c \ox (\proj{0}\ox U_B U_A+ \proj{1} \ox U_A U_B)\ket{\phi} \nonumber   \nonumber 
\\
&+\ket{6_A,6_B}_c \ox \mathcal{G}(U_A,U_B)\ket{\phi}+\ket{7_A,7_B}_c \ox \mathcal{G}(U_A,U_B)\ket{\phi}, \nonumber
\end{align}
where $\mathcal{G}(U_A,U_B)=\proj{0}\ox U_B U_A+ \proj{1} \ox U_A U_B$ is the (pure) process matrix. Intuitively the history state in Equation \eqref{historySWITCH} describes the scenario where, depending on the value of the control, different time orderings ($A$'s clock ticks at a faster rate than $B$'s or vice versa) are initiated by de-synchronizing initially synchronized clocks. For the two time orderings different orders of the agents' operations (either $U_A$ or $U_B$ first) are applied. Finally the clocks are re-synchronized, again making use of the control degree of freedom, such that they can tick together at the end of the protocol. We obtain the following perspectival states

\begin{align}
&\ket{\psi_A(0)}=\ket{0_B}_{c_B}\ox \ket{\phi} &&\ket{\psi_B(0)}=\ket{0_A}_{c_A} \ox \ket{\phi} \nonumber \\
&\ket{\psi_A(1)}=\ket{1_B}_{c_B}\ox \ket{\phi} &&\ket{\psi_B(1)}=\ket{1_A}_{c_A}\ox \ket{\phi} \nonumber\\
&\ket{\psi_A(2)}=\ket{2_B}_{c_B}\ox  (\proj{0}\otimes \mathds{1})\ket{\phi}
&&\ket{\psi_B(2)}=\ket{2_A}_{c_A}\ox  (\proj{1}\otimes \mathds{1})\ket{\phi} \nonumber \\
&\qquad\qquad+\frac{1}{\sqrt{2}}(\ket{2_B}+\ket{3_B})_{c_B}\ox (\proj{1}\otimes \mathds{1})\ket{\phi}
&&\qquad\qquad+\frac{1}{\sqrt{2}}(\ket{2_A}+\ket{3_A})_{c_A}\ox (\proj{0}\otimes \mathds{1})\ket{\phi}\nonumber \\
&\text{with }N_2^{(A)}=\proj{0}_{Sc}+\frac{1}{\sqrt{2}}\proj{1}_{Sc}
&&\text{with }N_2^{(B)}=\frac{1}{\sqrt{2}}\proj{0}_{Sc}+\proj{1}_{Sc}\nonumber\\
&\ket{\psi_A(3)}=\ket{2_B}_{c_B}\ox  (\proj{0}\otimes \mathds{1})\ket{\phi}
&&\ket{\psi_B(3)}=\ket{2_A}_{c_A}\ox  (\proj{1}\otimes \mathds{1})\ket{\phi} \label{Equation:SwitchPerspectives}\\
& \qquad\qquad+\ket{4_B}_{c_B}\ox  (\proj{1}\otimes U_B)\ket{\phi}
&& \qquad\qquad+\ket{4_A}_{c_A}\ox  (\proj{0}\otimes U_A)\ket{\phi} \nonumber
\end{align}
\begin{align}
&\ket{\psi_A(4)}=\ket{3_B}_{c_B}\ox  (\proj{0}\otimes U_A)\ket{\phi}
&&\ket{\psi_B(4)}=\ket{3_A}_{c_A}\ox  (\proj{1}\otimes U_B)\ket{\phi} \nonumber \\
& \qquad\qquad+\ket{5_B}_{c_B}\ox  (\proj{1}\otimes U_A U_B)\ket{\phi}
&& \qquad\qquad+\ket{5_A}_{c_A}\ox  (\proj{0}\otimes U_B U_A)\ket{\phi} \nonumber\\
&\ket{\psi_A(5)}=\frac{1}{\sqrt{2}}(\ket{4_B}+\ket{5_B})_{c_B}\ox  (\proj{0}\otimes U_B U_A)\ket{\phi}
&&\ket{\psi_B(5)}=\frac{1}{\sqrt{2}}(\ket{4_A}+\ket{5_A})_{c_A}\ox  (\proj{1}\otimes U_A U_B)\ket{\phi} \nonumber \\
& \qquad\qquad+\ket{5_B}_{c_B}\ox  (\proj{1}\otimes U_A U_B)\ket{\phi}
&& \qquad\qquad+\ket{5_A}_{c_A}\ox  (\proj{0}\otimes U_B U_A)\ket{\phi} \nonumber\\
&\text{with }N_5^{(A)}=\frac{1}{\sqrt{2}}\proj{0}_{Sc}+\proj{1}_{Sc} 
&&\text{with }N_5^{(B)}=\proj{0}_{Sc} +\frac{1}{\sqrt{2}}\proj{1}_{Sc} \nonumber\\
&\ket{\psi_A(6)}=\ket{6_B}_{c_B}\ox\mathcal{G}(U_A,U_B) \ket{\phi} &&\ket{\psi_B(6)}=\ket{6_A}_{c_A}\ox\mathcal{G}(U_A,U_B) \ket{\phi} \nonumber \\
&\ket{\psi_A(7)}=\ket{7_B}_{c_B}\ox\mathcal{G}(U_A,U_B) \ket{\phi} &&\ket{\psi_B(7)}=\ket{7_A}_{c_A}\ox\mathcal{G}(U_A,U_B) \ket{\phi}\nonumber
\end{align}
which can be related to each other by unitaries
\begin{align}
&\mathcal{U}_A(1,0)=T_{c_B} \ox \mathds{1}_S &&\mathcal{U}_B(1,0)=T_{c_A}\ox \mathds{1}_S \nonumber\\
&\mathcal{U}_A(2,1)= T_{c_B}\ox (\proj{0} \ox \mathds{1})_S+(T'_{2})_{c_B}\ox (\proj{1} \ox \mathds{1})_S &&\mathcal{U}_B(2,1)= T_{c_A}\ox (\proj{1} \ox \mathds{1})_S+(T'_{2})_{c_A}\ox (\proj{0} \ox \mathds{1})_S \nonumber\\
&\mathcal{U}_A(3,2)=  \mathds{1}_{c_B}\ox (\proj{0} \ox \mathds{1})_S+(T'_{2})_{c_B}\ox (\proj{1} \ox U_B)_S
&&\mathcal{U}_B(3,2)= \mathds{1}_{c_A}\ox (\proj{1} \ox \mathds{1})+(T'_{2})_{c_A}\ox (\proj{0} \ox U_A) \nonumber\\
&\mathcal{U}_A(4,3)= T_{c_B}\ox(\mathds{1}\ox U_A)_S&&\mathcal{U}_B(4,3)= T_{c_A}\ox(\mathds{1} \ox U_B)_S \label{Equations:SwitchUnitaries}\\
&\mathcal{U}_A(5,4)= (T'_{4})_{c_B}\ox (\proj{0} \ox U_B)_S+\mathds{1}_{c_B}\ox (\proj{1} \ox \mathds{1})_S
&&\mathcal{U}_B(5,4)=(T'_{4})_{c_A}\ox(\proj{1} \ox U_A)_S+\mathds{1}_{c_A}\ox (\proj{0} \ox \mathds{1})_S \nonumber\\
&\mathcal{U}_A(6,5)= (T'_{4})_{c_B}\ox (\proj{0} \ox \mathds{1})_S+ T_{c_B}\ox (\proj{1} \ox \mathds{1})_S
&&\mathcal{U}_B(6,5)= (T'_{4})_{c_A}\ox (\proj{1} \ox \mathds{1})_S+ T_{c_A}\ox (\proj{0} \ox \mathds{1})_S \nonumber\\
&\mathcal{U}_A(7,6)=T_{c_B}\ox \mathds{1}_S &&\mathcal{U}_B(7,6)=T_{c_A}\ox \mathds{1}_S.
\nonumber
\end{align} 
where $T$ and $T'_{i}$ are the same as in the previous example. It is straightforward to see that all our axioms are fulfilled. Note that the unitaries in Eqs.(\ref{Equations:SwitchUnitaries}) are not unique but were chosen such that the perspectives of the agents resemble the causal reference frames of the two agents presented in Ref.~\cite{guerin2018agent}. For both $A$ and $B$ the time of action is $t^{\star}_A=4=t^{\star}_B$ and depending on the value of the control the other agent applies their unitary either before or after $t^{\star}_A$ or $t^{\star}_B$ respectively, compare to Fig.\ref{Fig:QSwitch}.

%=====================================%

\subsection{General coherent control of causal order}
\label{Section:ControlledCombs}

%In the previous section we have seen how to implement the quantum switch in our framework. The quantum switch is a famous example for an important class of processes with indefinite causal structure: The processes with coherently controlled causal order. These are processes controlled by a quantum degree of freedom. To each value of the control system one associates a process with definite causal order \textcolor{blue}{(a quantum comb, as we describe below)}, but this definite causal order can be different for different control values. In particular, as the control system is a quantum system, one can have controlled superpositions of different causal orders, just as we have seen for the quantum switch.\\

The quantum switch from the previous section is a famous example for an important class of processes with indefinite causal structure, namely processes with coherently controlled causal order~\cite{PurvesQuantumControlledOrder, WechsQuantumControlledOrder}. There, for each value of the control system $\ket{k}\in \mathcal{H}_{Sc}$, one associates a process with definite causal order or quantum comb $\mathcal{\tilde G}_k$~\cite{CombsLong, CombsPRA} and the definite causal order is different for at least two different $k$ {\footnote{We do not consider the more general case in which one agent can control the order of the other agents.}. We will now present the general idea for implementing such processes in our framework, while the mathematical details are given in Appendix~\ref{Appendix:Combs}. \\

Consider $M$ pure combs $\mathcal{\tilde G}_k$, $1 \le k \le M$ controlled by an $M$-dimensional control system, with control state $\ket{k}_Sc$ meaning that $\mathcal{\tilde G}_k$ will be implemented. We will label the $N$ agents as $A_1,\dots, A_N$ and their unitary operations as $U_{1},\dots , U_N$. As explained in Ref.~\cite{TulioSuperchannels}, quantum combs can be represented by a sequence of channels with memory, see Fig.~\ref{Figure:MixedCombsAndDilation} for a tripartite example, and we can write processes with coherently controlled causal order as
\begin{align}
\mathcal{G}(U_{1}, \dots, U_{N})&=\sum_{k=1}^M \proj{k}_C \ox \mathcal{\tilde G}_k(U_{1}, \dots, U_{N}) \\
&=\sum_{k=1}^M \proj{k}_C \ox V^{(k)}_N U_{{\pi_k(N)}} V^{(k)}_{N-1} U_{{\pi_k(N-1)}} \dots V^{(k)}_1 U_{{\pi_k(1)}} V^{(k)}_0,
\label{Eq:CohCoCausal}
\end{align}
where the $V^{(k)}_0, \dots, V^{(k)}_N$ are unitaries corresponding to purified channels with memory, which act trivially on the ancillas $S' = A'_{1} \otimes \dots \otimes A'_N$ the agents use. The unitaries $U_{{\pi_k(1)}} \dots U_{{\pi_k(N)}}$ are a permutation of the agents' unitaries $U_1,\dots , U_N$, which represent the particular definite causal order realized in the comb $\mathcal{\tilde G}_k$. As one can see, $\mathcal G$ is unitary and multilinear in the operations of the agents. Such processes were also considered in~\cite{PurvesQuantumControlledOrder, WechsQuantumControlledOrder}.

\begin{figure}[h!]
\includegraphics[width = 0.8\textwidth]{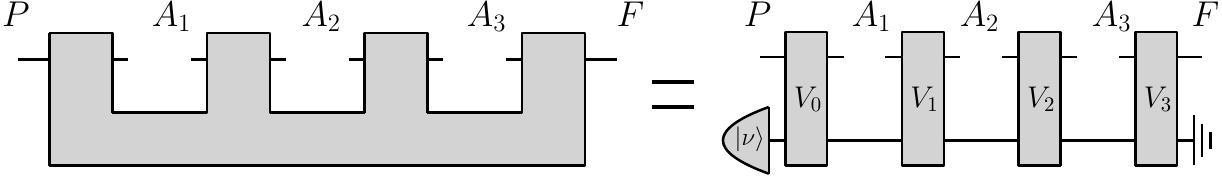}
\caption{A tripartite quantum comb: A general processes with fixed causal order is a map on the actions of three agents $A_1$,  $A_2$ and $A_3$ (left). Time passes from left to right, where $P$ stands for past and $F$ for future, and, hence, $A_1$ acts before $A_2$ and agent $A_3$ acts last. The agents' actions are quantum instruments which must be inserted into the slots of the comb. Every comb can be implemented as sequence of unitary channels with memory when adding an additional environment system with input $\ket{\nu}$ and discarding part of the output (right)~\cite{CombsLong,CombsPRA,Stinespring,TulioSuperchannels}. For pure combs no extra environment input or discarding operation is required~\cite{TulioSuperchannels}.}
\label{Figure:MixedCombsAndDilation}
\end{figure}

Now we describe the history state implementing $\mathcal{G}(U_{1}, \dots, U_{N})$ given by Eq.~\eqref{Eq:CohCoCausal}. The input state $\ket{\psi}_S \in \mathcal{H}_{Sc} \ox \mathcal{H}_{Sp}$ comprises a control system ($\in \mathcal{H}_{Sc}$) and another system ($\in \mathcal{H}_{Sp}$) which represent the input to the combs from the global past. We decompose the protocol and, hence, the history state into three parts as 
\begin{align}
	\ket{\Psi \rangle} = \ket{\Psi_{\text{desync}} \rangle} + \ket{\Psi_{\text{combs}} \rangle} + \ket{\Psi_{\text{resync}}\rangle},
\end{align}
where $\ket{\Psi_{\text{desync}} \rangle} $ describes the beginning of the protocol, where we use the control degree of freedom to desynchronize the clocks such that the agents are put into the right order. Afterwards the different combs are applied depending on the value of the control in $\ket{\Psi_{\text{combs}} \rangle}$. At last, the resynchronization of the agents' clocks is described by $\ket{\Psi_{\text{resync}}\rangle}$. The strategy is depicted in Fig.\ref{Figure:CoherentPlan2}.\\

\begin{figure}[h!]
\includegraphics[width = \textwidth]{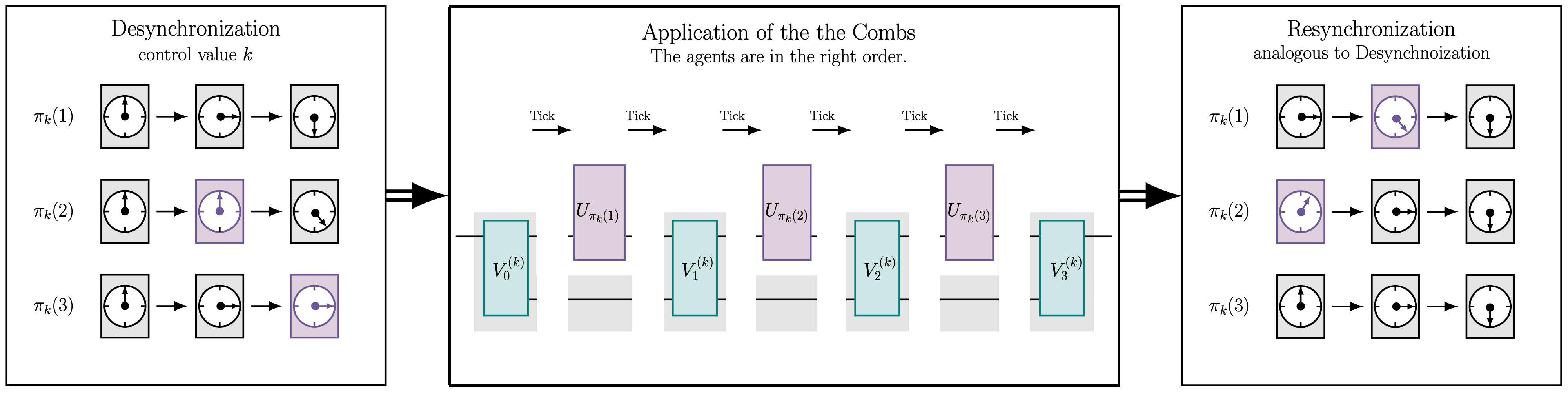}
\caption{The strategy for implementing coherently controlled causal order: The protocol consists of three steps which are all conditioned on the control value, namely desynchronization, application of the combs and resynchronization. The time of action $t^*$ is chosen the same for all agents. To be able to implement the comb of a given $k$, the clocks of the agents get desynchronized such that the agents will act in the right order. This is achieved by first freezing the clocks of all but the agent $A_{\pi_k(1)}$ one after the other. By "freezing" we mean that all readings of clocks other than that of agent $A_{\pi_k(1)}$ are not changing while the clock of agent $A_{\pi_k(1)}$ ticks. First the clock of the agent acting second gets frozen followed by that of the agent acting third etc. The clock freezes are indicated in purple. The duration of the freeze depends on when the agent will act. After the desired orderings have been implemented, the combs get applied while all clocks tick in synchronization. First one applies the first unitary with memory of the comb, then the first agent acts (their clock shows $t^*$). Then the second comb unitary is applied followed by second agents' action etc. At last the clocks get resynchronized again by using the desynchronization protocol, but with the role of the agents reversed.}
\label{Figure:CoherentPlan2}
\end{figure}

During the desynchronization of the clocks nothing happens to the input to the combs and we can write
\begin{align}
\ket{\Psi_{\text{desync}} \rangle} =& \ket{0,\dots,0}_{c} \ox \ket{\psi}_S +  \ket{1,\dots,1}_{c} \ox \ket{\psi}_S+ \ket{2,\dots,2}_{c} \ox \ket{\psi}_S \\
	&+   \sum_{k=1}^M \sum_{j=3}^{T_0} \ket{t^{(k)}_{1}(j), \dots, t^{(k)}_{N}(j)}_c \ox (\proj{k}_{Sc} \otimes \mathbb 1_{Sp}) \ket{\psi}_S , \nonumber
\end{align}
where, as we will see, $T_0 := t^* - 2$, with $t^*$ being the time of action for all agents. The $t^{(k)}_i(j)$ give different time orderings by freezing different clocks for different amounts of time steps during which the other clocks keep ticking. More precisely, if an agent will act as the $m$-th agent in the comb with the control value $\ket{k}_{S_C}$, then the clock of that agent gets frozen for $2(m-1)$ time steps. This ensures that two consecutive agents are two time steps apart from each other when they enter $\kket{\Psi_{\text{combs}}}$. While one agent's clock is frozen, the clocks of the other agents march on. See Appendix~\ref{Appendix:Combs} for the detailed clock freezing and desynchronization protocol. Afterwards, we include additional synchronized ticks at the end of the desynchronization step to ensure that for all $k$ the clock freezes are far away from the application of the combs. This together with the fact that the agents' operations $U_j$ have not been used, yet, gives perspectival, controlled unitaries that act non-trivially only on the clocks of the other agents, i.e.  $\mathcal U_{A_j}(t,t-1) = \sum_{k=1}^M u^{A_j}_{c,k}(t,t-1)\ox \proj{k}_{Sc} \otimes \mathbb 1_{Sp} $. Here, $u^{A_j}_{c,k}(t,t-1)$ are the unitaries that only act on the clocks of the other agents.\\

In $\kket{ \Psi_{\text{combs}}}$ the clocks will continue to tick in synchronization by means of the unitary $T$ introduced in Section~\ref{"Twin-paradox" circuit} while, given a control value $k$, the unitaries of the comb $\mathcal{\tilde G}_k$ are applied one after the other. All the agents, for a given $k$, see the following sequence of unitaries at the respective time steps:

\begin{align}
	V^{(k)}_0 \otimes T^{\otimes (N-1)},\ U_{{\pi_k(1)}}\otimes T^{\otimes (N-1)},\  V^{(k)}_1\otimes T^{\otimes (N-1)}, \ U_{{\pi_k(2)}}\otimes T^{\otimes (N-1)}, \ \dots, \ U_{{\pi_k(N)}}\otimes T^{\otimes (N-1)}, \ V^{(k)}_N \otimes T^{\otimes (N-1)}
\end{align}

Further details are again in Appendix \ref{Appendix:Combs}. The time differences caused by freezing the clocks ensure that the time of action $t^*$ satisfies $t^* = T_0 +2$ for all agents. \\

For the resynchronization in $\ket{\Psi_{\text{resync}} \rangle}$ we repeat the procedure from $\ket{\Psi_{\text{desync}} \rangle}$, but with the role of the agents inverted, i.e. $t^{(k)}_{\pi_k(m)} \mapsto t^{(k)}_{\pi_k(N+1-m)}$. In the end, all the clocks tick in synchronization and show the same time. Like the desynchronization this last part of the protocol is independent of the agents' operations $U_j$ and our axioms are fulfilled. Hence, any coherent control of causal order as described by Eq.~\eqref{Eq:CohCoCausal} can be implemented in our framework.

%=====================================%
\subsection{About an exotic process}
\label{sec:Lugano process}

 A notorious example of a tripartite pure process with indefinite causal order from~\cite{araujo2017purification,baumeler2016space} is known to violate causal inequalities. Said process is \emph{not} an example of coherent control of causal order. It is often referred to as the \emph{Lugano process}. The time reversed version of the Lugano process was discussed in Ref.~\citep{guerin2018agent} and can be written as 
\begin{align}
&\mathcal{G}(U_A,U_B,U_C)\ket{jjj}=U_A \ox U_B \ox U_C \ket{jjj} \label{eqLugano1}\\
&\mathcal{G}(U_A,U_B,U_C)\ket{j01}=XU_A \ox U_B \ox U_C \ket{j01} \\
&\mathcal{G}(U_A,U_B,U_C)\ket{1j0}=U_A \ox XU_B \ox U_C \ket{1j0} \\
&\mathcal{G}(U_A,U_B,U_C)\ket{01j}=U_A \ox U_B \ox XU_C \ket{01j} \label{eqLugano4}
\end{align}
where $j\in \{ 0,1\}$ and $X = \sigma_X$ is the Pauli-$X$ matrix. Defining projectors $P_A=\sum_j \proj{j01} $, $P_B=\sum_j \proj{1j0}$, $P_C=\sum_j \proj{01j}$ and $P_{\perp}=\sum_j \proj{jjj}$ one gets
\begin{align}
&\mathcal{G}(U_A,U_B,U_C)\ket{\phi}=
(U_A \ox U_B \ox U_C P_{\perp}
+XU_A \ox U_B \ox U_C P_A
+U_A \ox XU_B \ox U_C P_B
+U_A \ox U_B \ox XU_C P_C)\ket{\phi}.
\label{Equation:LuganoWithProjectors}
\end{align}

%As the reversed Lugano process violates causal inequalities, it cannot be implemented via a usual circuit with clocks that are constantly well-synchronized. Therefore, it is necessary to desynchronize the clocks in some way. 
One crucial difference between the reversed Lugano process and the non-causal processes discussed in Section~\ref{Section:ControlledCombs} is the lack of a control degree of freedom. Therefore, it is not possible to directly adapt the history state procedure that we used for coherently controlled causal order to the reversed Lugano process. Instead, the main system itself has to control the desynchronization process. %The decomposition of Equation (\ref{Equation:LuganoWithProjectors}) suggests one way how to do this, but as we will argue also this way will fail:
One can try to, similarly to the quantum switch, use the projectors $P_A$,$P_B$,$P_C$ and $P_{\perp}$ to define a controlled operation that de-synchronizes the clocks.  Afterwards, one can use the clocks as a control system to define another controlled operation that applies the unitary operations~\eqref{eqLugano1}- \eqref{eqLugano4} for the different control values. However, the re-synchronization cannot be done independently of the unitaries $U_A$, $U_B$ and $U_C$. 
More specifically, the described procedure will lead to a term in the history state of the form
\begin{align*}
	\ket{\Psi \rangle} = \dots +  \ket{\gamma_\perp}_c \ox ( U_A \ox U_B \ox U_C P_{\perp})\ket{\phi}_S  + \ket{\gamma_A}_c \ox (XU_A \ox U_B \ox U_C P_A)\ket{\phi}_S \\
	+ \ket{\gamma_B}_c \ox (U _A \ox XU_B \ox U_C P_B)\ket{\phi}_S + \ket{\gamma_C}_c \ox (U_A \ox U_B \ox XU_C P_C)\ket{\phi}_S + \dots 
\end{align*}
with some clock states $\ket{\gamma_\perp}_c$, $\ket{\gamma_A}_c$, $\ket{\gamma_B}_c$ and $\ket{\gamma_C}_c$, which represent the different time orderings. The question is how to complete the history state, i.e. how to resynchronize the clocks. 
We are only allowed to use the agents' operations once and this has already happened. The states $U_A \ox U_B \ox U_C P_{\perp} \ket{\phi}_S$, $XU_A \ox U_B \ox U_C P_A\ket{\phi}_S$, $U_A \ox XU_B \ox U_C P_B\ket{\phi}_S$ and $U_A \ox U_B \ox XU_C P_C\ket{\phi}_S$ all depend on $U_A,U_B,U_C$ in different, non-trivial ways. This means any overall map using them to ``resynchronize'' the clocks will non-trivially depend on $U_A,U_B$ and $U_C$ as well. This in turn leads to a non-trivial dependence of $\UU_X(t_X,t_X-1)$ on $U_X$ for all $X \in \{A,B,C \}$ during the resynchronization part towards the end of the protocol, i.e. for $t_X>t_X^*$, which is a violation of Assumption {\bf U.3}.\\

\begin{figure}[h!]
%\vspace{-1em}
\includegraphics[width=0.75\linewidth]{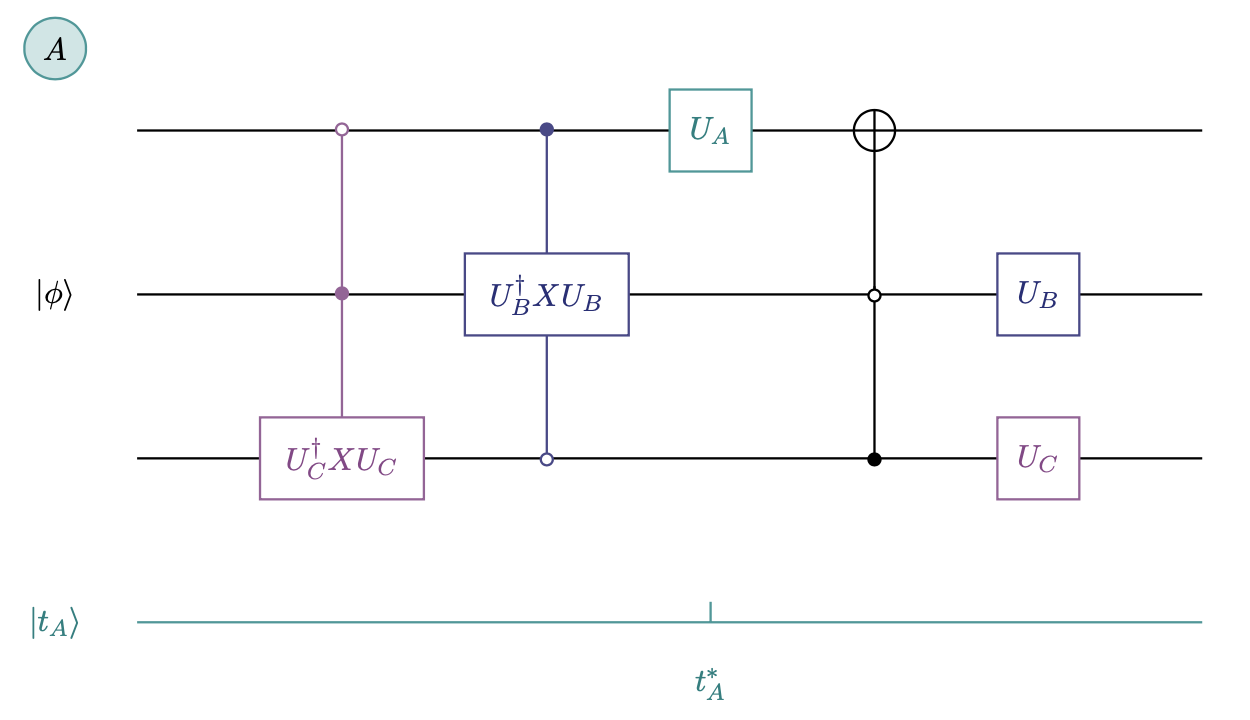} 
\caption{The causal reference frame of agent $A$ inside the time reversed Lugano process as given in~\cite{guerin2018agent} ($B$'s and $C$'s perspectival circuits look analogous). There is no control degree of freedom. All three agents act on different subsystems of the input system $\ket{\phi}$, but in a way that depends on the subsystems the other agents act on. Because of the gates that are not affine-linear in $U_B$ and $U_C$ this causal reference frame decomposition is incompatible with our setting.
}
\label{Fig:Lugano}
\end{figure}

Ref.~\cite{guerin2018agent} presented a causal reference frame decomposition of the reverse Lugano process, which for agent $A$ is shown in Fig.~\ref{Fig:Lugano}. It uses perspectival circuits with gates that are not affine-linear in the respective unitaries of the other agents, $U_B^\dagger X U_B$ and $U_C^\dagger X U_C$ for $A$'s perspective. However, the corresponding perspectival states are forbidden in our framework due to the requirement of affine-linearity for $M_{t_A,t_B,t_C}(U_A,U_B,U_C)$ discussed in Section \ref{Section:AffineLinearM}.\\

Note, however, that the two impossible implementations of the process $\mathcal{G}(U_A,U_B,U_C)$  discussed above, namely the causal reference frame decomposition of \cite{guerin2018agent} according to Fig.~\ref{Fig:Lugano} and the desynchronization-resynchronization-protocol, are not necessarily the only strategies for how to describe the reverse Lugano process within our non-causal Page-Wootters framework. Determining whether this process can be realized in this framework remains an open problem left for future work.

%%%%%%%%%%%%%%%%%%%%%%%%%%%%%%%%%%%%%%%%%%%%%%%%%%%%%%%

\section{Conclusion}
\label{Discussion}

In this paper we showed how the Page-Wootters approach and the process matrix formalism may be combined to give a history state description of non-causal processes. We considered an operational setting that allows for probing indefinite causal structure. In this setting we explicitly modeled the passage of time as perceived by different agents using discrete quantum clocks. This allowed us to use a history state approach to which we added a set of well-motivated axioms about the protocol and the perspectives of the agents. As a consequence of these axioms, the causal structures arising in our setting are described by pure process matrices. A well-known result from previous literature about pure process matrices implies that in the bipartite case, no violation of device-independent causal inequalities can occur in our setting. Nonetheless, we could show that important physical scenarios beyond causal circuits and beyond non-relativistic clocks fit into our framework. More specifically, we showed how to describe a scenario inspired by the twin paradox involving varying clock ticking speeds with our approach. But most importantly, we proved that all processes representing coherent control of causal order (e.g. the quantum switch) can be implemented using our description.\\

We showed how to extract the time-evolution corresponding to the perspective of any given agent. This  lead us to a refinement of the \emph{causal reference frame} picture of Ref.~\cite{guerin2018agent} in which also the quantum clocks are explicitly modeled. The presence of these clocks and a perspective-neutral history state impose extra conditions on the causal reference frames. As an example, we showed that the  evolution described by the causal past needs to be affine-linear in the operations of the other agents. We applied this extra condition to rule out, within our setting, a specific causal reference decomposition of the so-called time-reversed Lugano process provided in \cite{guerin2018agent}. \\

We conclude by pointing out a few directions for future research.  While we focused on discrete clocks, this framework can be adapted to continuous clocks, extending the approach of Ref.~\cite{Castro-Ruiz2020} to a systematic operational protocol that allows for the extraction of process matrices. In order to model the protocol for probing causal structure, we worked directly with history states instead of starting with a constraint operator or physical projector. As a consequence, the relation between the physical projector and the perspectival unitaries $\UU_X(t', t)$ is an open question. Resolving this question might reveal further constraints on the history states, possibly restricting the set of process matrices that can be described by our framework. An important class of non-causal processes that lack a physical interpretation are those that violate causal inequalities (e.g. the aforementioned Lugano process). If one could show that these processes do not fit into our setting, this would hint at such processes not being physical.

%We saw that the process matrices compatible with our setting are always pure. For the bipartite case this implies that no violation of causal inequalities can occur in our setting.\\

%Furthermore, we also saw that our framework leads to a modification of the \emph{causal reference frame} picture of \cite{guerin2018agent} in which also the quantum clocks are explicitly modeled. This imposed extra conditions on the causal reference frame. These conditions can be interpreted as compatibility requirements on how one agent can perceive the actions of the other agents. These extra conditions allowed us to rule out a causal reference frame decomposition presented in \cite{guerin2018agent} for an important exotic process, the time-reversed \emph{Lugano process}.\\

%As specific examples we showed how two important scenarios fit into our setting: The twin paradox involving varying clock speeds  and the quantum switch which is non-causal.\\
 
%Analyzing process matrices within a history states approach allows to make progress on the problem that most process matrices lack a physical interpretation. For example, we presented an operational setup to describe the probing of causal structure. One obtains a more refined description of the involved physics by explicitly modeling the quantum clocks of the agents. This induces extra compatibility conditions between the perspectives of the agents that in turn allow to exclude some process matrices. \\

\section{Acknowledgements}
We thank Esteban Castro-Ruiz, Marco T\'{u}lio Quintino and David Trillo Fernandez for interesting discussions. We acknowledge the support of the Vienna Doctoral School in Physics (VDSP) and the support of the Austrian Science Fund (FWF) through the Doctoral Programme CoQuS. \v{C}.B.  acknowledges financial support from the Austrian Science Fund (FWF) through BeyondC (F7103-N48), from the European Commission via Testing the Large-Scale Limit of Quantum  Mechanics  (TEQ)  (No.  766900)  project,  and from  Foundational Questions Institute (FQXi). This research was supported by FQXi FFF Grant number FQXi-RFP-1815 from the Foundational Questions Institute and Fetzer Franklin Fund, a donor advised fund of Silicon Valley Community Foundation.  We acknowledge a grant from the John Templeton Foundation (ID\#~61466) as part of the The Quantum Information Structure of Spacetime (QISS) Project (qiss.fr). Research at Perimeter Institute is supported in part by the Government of Canada through the Department of Innovation, Science and Economic Development Canada and by the Province of Ontario through the Ministry of Colleges and Universities.

%%%%%%%%%%%%%%%%%%%%%%%%%%%%%%%%%%%%%%%%%%%%%%%%%%%%%%%
%\bibliographystyle{unsrt}
%\bibliographystyle{apsrev4-2}
%\bibliography{page_wooters}

%apsrev4-2.bst 2019-01-14 (MD) hand-edited version of apsrev4-1.bst
%Control: key (0)
%Control: author (8) initials jnrlst
%Control: editor formatted (1) identically to author
%Control: production of article title (0) allowed
%Control: page (0) single
%Control: year (1) truncated
%Control: production of eprint (0) enabled
%

\newpage
%%%%%%%%%%%%%%%%%%%%%%%%%%%%%%%%%%%%%%%%%%%%%%%%%%%%%%%

\appendix

\section{Discretization and normalization operators}
\label{Appendix:Normalization}

%In this section we will investigate the issue of discretization and the use of normalization operators in more detail. \\

%In relativity, the proper time associated to a worldine is a continuous variable, so it is natural to associate a continuous Hilbert space to the agents' clocks, as was done in Ref.~\cite{Castro-Ruiz2020}. However, to make the connection with the circuit formalism easier, we chose instead to use discrete clocks. This change from continuous to discrete time introduces two issues that can be tackled with the introduction of normalization operators. \\
An important difference between continuous and discrete clocks is the fact that integrals $\int \mathrm d t$ pick up prefactors when changing the integration variable while sums do not pick up such a prefactor under change of summation index. Consider again the example from the main text of the history state $\ket{\Psi \rangle} =\int \mathrm dt_A \ket{t_A}_{c_A} \otimes \ket{2t_A}_{c_B}$ with perspectival states $_{c_A}\braket{ t_A }{ \Psi \rangle} = \ket{2t_A}$ and $_{c_B}\braket{t_B}{\Psi \rangle} = \int \mathrm{d} t_A \ket{t_A} \braket{t_B}{2t_A} = \frac{1}{2} \ket{1/2 \ t_B}$.
If we consider a naive discretization of the above example of the form $\ket{\Psi \rangle} = \sum_k \ket{k}_{c_A} \otimes \ket{2k}_{c_B}$ we find $_{c_A}\braket{t_A}{\Psi \rangle} = \ket{2 t_A}$ and for even values of $t_B$ we find $_{c_B}\braket{t_B}{\Psi \rangle} =\ket{1/2\ t_B}$. Hence, the continuous and discrete version of $_{c_B}\braket{t_B}{\Psi \rangle}$ differ by a factor $\frac 1 2$. \\

The second important issue arises from the use of approximations like time-binning, i.e. to assign every continuous time state $\ket{t}$ to the closest discrete time state, for discretization. Such procedures are not injective: Continuous times $\ket{t+\delta t}$ and $\ket{t}$ with very small $\delta t$ will in general get mapped to the same discrete time state. This means that the discretization procedure itself can change the normalization and inner product of states. Such artifacts of the discretization procedure can be countered by the introduction of normalization operators. \\

For well-synchronized clocks with constant and same ticking speed, the aforementioned discretization artifacts can usually be avoided. However, we will show now issues one encounters in the context of different or varying clock ticking speeds. As a warm-up, let us consider again a history state of a clock that ticks twice as fast as another clock: $\int \mathrm{d} t \ket{t}_{c_A} \otimes \ket{2t}_{c_B}$. A first guess for a discretization might be something of the form $|\Psi \rrangle = \sum_{k} \ket{k}_{c_A}\otimes \ket{2k}_{c_B}$ with $k$ taking integer values. This would not be an acceptable discretization in our approach: Our postulates demand that agent $B$ sees a state for each time value of their clock, and that this state evolves via unitary time evolution. However, $_{c_B}\langle t|\Psi \rrangle = 0$ for $t$ odd makes this impossible.

One possible approach to fix this issue might be to instead use $|\Psi \rrangle = \sum_{k} \ket{k}_{c_A}\otimes \ket{k}_{c_B}$ and keep a note that says that the times on $B$'s clock must be multiplied by another factor of $2$ to obtain the ``real'' time of Bob. Such a fix is very unappealing and goes against the idea that the clock states directly reflect the time of the agents, up to rounding error. Also, in the context of varying ticking rates the implementation of such a fixing strategy can become very complicated. The situation becomes even worse in the context of superpositions of histories: Here, the mapping of $\ket{k}_{c_B}$ to the actual value of Bob's clock might depend on the branch of the superposition and the same $\ket{k}_{c_B}$ might correspond to vastly different times on $B$'s clock. An example might be a history state that is a superposition of $A$'s clock being twice as fast and and $B$'s clock being twice as fast, i.e. $\int \mathrm{d} t (\alpha \ket{2t} \otimes \ket{t} + \beta \ket{t} \otimes \ket{2t})$. Obviously, now a ``fix'' like $\sum_k (\alpha \ket{k}\otimes\ket{k} + \beta \ket{k}\otimes\ket{k})$ cannot work.

Let us look for a good history state that can describe discrete clocks of different ticking rates. We would like the clock states to directly tell us the time of the clock. Also, as we argued before, our discretization of clocks is not allowed to leave out any times. Then, to describe clocks of different speeds, one is left with the option to instead repeat times: A discretization of $\int \mathrm d t \ket{2t}_{c_A} \otimes \ket{t}_{c_B}$ might be $|\Psi \rrangle = \sum_k \ket{k}_{c_A} \otimes \ket{\lfloor \frac{k}{2} \rfloor}_{c_B}$, with $\lfloor \bullet \rfloor$ meaning ``rounded down''. As a specific example, let us consider the state
\begin{align}
 	\ket{\Psi \rangle} = \ket{0} \otimes \ket{0} +\ket{1} \otimes \ket{1} + \ket{2} \otimes \ket{1} + \ket{3} \otimes \ket{2} +\ket{4} \otimes \ket{2} + \ket{5} \otimes \ket{3}\ldots 
\end{align} 
Note that, this state does not satisfy all our axioms for history states from the main text and it is intended as an illustration; all degrees of freedom other than the clocks have been neglected. This procedure of repeating times does have a nice interpretation: One can interpret the clock states $\ket{k}$ as the number of ticks the agent has heard so far. As $A$'s clock is twice as fast, $B$ hears the first tick when $A$ already hears the second.\\
 
Let us see what the states for the different perspectives look like. We have 
\begin{align*}
&_{c_A}\langle 0 |\Psi \rrangle = \ket{0}_{c_B}, & &_{c_A}\langle 1 |\Psi \rrangle  = {\ket{1}_{c_B}}, & &_{c_A}\langle 2 |\Psi \rrangle = \ket{1}_{c_B}, & &_{c_A}\langle 3 |\Psi \rrangle  = {\ket{2}_{c_B}}, & &_{c_A}\langle 4 |\Psi \rrangle = \ket{2}_{c_B}, && \ldots
\end{align*}
This fits to the interpretation that whenever $B$ hears one tick, $A$ already hears the second tick. Note that the states are properly normalized. For $B$'s perspective we find
\begin{align*}
	&_{c_B}\langle 0|\Psi \rrangle  = { \ket{0}_{c_A}}, && _{c_B}\langle 1 |\Psi \rrangle = { \ket{1}_{c_A} + \ket{2}_{c_A}}, && _{c_B}\langle 2 |\Psi \rrangle  ={ \ket{3}_{c_A} + \ket{4}_{c_A}}, && \ldots
\end{align*}
First we note that these states are not properly normalized. But this can be easily fixed with a normalization operator $N^{(B)}_{t_B} = \frac{1}{\sqrt 2} \mathbb 1$ for $t_b > 0$. Indeed, this normalization factor arises because we map continuous times $\ket{k+\Delta t}_{c_B}$,with $k$ an integer, $0 \le \Delta t < 1$, to the same discrete state $\ket{k}_{c_B}$, as mentioned previously. Furthermore we note that $B$ ``coherently interpolates'' between the two times of $A$ that are consistent with $B$'s time. Also this is reasonable: As $B$ cannot have ``which-time''-information about $A$'s clock without a measurement (in analogy to which-path-information), $B$ puts the two possible times in superposition.
%============================================================================================%

\section{About the physical projector and its relation to unitary time evolution}
\label{App:P}

It is unclear whether the unitaries $\UU_X(t'_X,t_X)$ can always be chosen such that they satisfy a nice relationship, similar to Eq.\eqref{Equation:CircuitUfromP}, with $\hat{P}_H$. The examples considered in Ref.~\cite{Castro-Ruiz2020} would seem to suggest an equation of the form
\begin{align}
\bra{t'_X} \hat{P}_H \ket{t_X} &\stackrel{?}{=} (N_{t'}^{(X)})^{-1} \UU_X(t_X', t_X) N_t^{(X)}. \label{Equation:ProjectorAndUnitaries}
\end{align}
However, we will show now that there exist choices of history states $\ket{\Psi \rangle}$ and time evolutions $\UU_X(t'_X,t_X)$ that are compatible with our framework as presented in Section \ref{Section:OurFormalism}, but do not satisfy Eq.~\eqref{Equation:ProjectorAndUnitaries}.\\

If Eq.~\eqref{Equation:ProjectorAndUnitaries} was true we could alternatively write
\begin{equation}
\hat P = \sum_{t, t'} |t' \rangle \langle t|_{T_A} \otimes (N_{t'}^{(A)})^{-1} \UU_A(t', t) N_t^{(A)},
\end{equation}
and a similar decomposition held for all other agents. Looking at two different ways to write out $\langle t_A'| \langle t_B'| P |t_A \rangle |t_B \rangle$, we have
\begin{equation}
\label{eq:consistency}
\langle t_B ' | N_A^{-1}(t_A') \UU_A (t_A', t_A) N_A(t_A) | t_B \rangle=\langle t_A ' | N_B^{-1}(t_B') \UU_B (t_B', t_B) N_B(t_B) | t_A \rangle. 
\end{equation}
By explicitly plugging in normalization operators $N_X$ and unitaries $\UU_X$ from example in Section~\ref{sec:switch} we can see that Eq.~\eqref{eq:consistency} does not hold for this representation of the quantum switch. More specifically, taking $t_B' = 3$, $t_B = 2$, $t_A' = 5$, $t_A = 4$, we obtain
\begin{equation}
\langle t_B=3|  N_A^{-1}(5) \UU_A (5, 4) N_A(4) | t_B = 2 \rangle= \sqrt{2} ( \langle 3| T_{4}' |2\rangle ) |0\rangle \langle 0| \otimes U_B = \sqrt{2} |0 \rangle \langle 0| \otimes U_B,
\end{equation}
which is not equal to
\begin{equation}
 \langle t_A=5 | N_B^{-1}(3) \UU_B (3, 2) N_B(2) | t_A =4 \rangle = \frac{1}{ \sqrt{2}} ( \langle 5 | T_{2}' |4\rangle )|0 \rangle \langle 0 | \otimes U_A = \frac{1}{\sqrt{2}} |0 \rangle \langle 0| \otimes U_A.
\end{equation}

Here, we extended $T'_i$ from the main text to act like the clock ticking operator $T$, when applied to $\ket{j}$ with $j \ne i-1, i, i+1$.

This example, however, does not mean that it is impossible to satisfy Eq.~\eqref{Equation:ProjectorAndUnitaries}. In our operational setting, we assumed that the ancillas and clocks are initialized to the states $\ket{0}$. Therefore, the states that emerge during the protocol do not probe the full input space of the $\UU_X(t'_X, t_X)$. In other words, there are several choices for $\UU_X(t'_X, t_X)$ that are compatible with our assumptions in Section \ref{Section:OurFormalism}. We leave for future work the question of whether there exists a choice of $\UU_X(t'_X, t_X)$ that simultaneously satisfies our postulates  and a relation similar to Eq.~\eqref{Equation:ProjectorAndUnitaries}.

%==========================================================================================%
\section{Coherently controlled causal order for an arbitrary number of agents}
\label{Appendix:Combs}

In this appendix we show explicitly how to implement coherently controlled causal order for an arbitrary number of agents. We start with some general considerations concerning the coherent superposition of quantum combs. Afterwards we present each of the three conceptual steps of the implementation, i.e. desynchronization, application of the combs and resynchronization, in detail.\\

First of all, in order to put quantum combs in controlled superposition they have to satisfy compatibility conditions. We assume that the input space and output space of an agent should be independent of the comb index $k$. Otherwise, an agent could narrow down the control value by determining the input or output dimension. Likewise, the dimension of the main system $S$, i.e. the dimension of the input to the comb from the global past and the dimension of the comb output to the global future, should be independent of the comb index $k$. Furthermore, we assume that the input and output space of an agent have the same dimension and that the memories of the combs are chosen such that their dimensions are independent of the comb index $k$. If the original combs do not satisfy these assumptions, the dimensions can be extended (for example by the use of ancillas) such that afterwards the combs dimensions do satisfy these requirements. \\ 

In order to implement quantum combs and their superpositions in our framework we use the fact that they can be modeled by sequences of unitary channels with memory, see Fig.~\ref{Figure:MixedCombsAndDilation} in the main text. More precisely, a general quantum comb $\mathcal{\tilde G}_k$ is given by a sequence of unitaries $V_0^{(k)}$,\dots, $V_N^{(k)}$ with memory and an environment input state $\ket{\nu^{(k)}}$ and an environment output system, over which the partial trace is taken at the end~\cite{TulioSuperchannels}. For our purpose we will consider an extended main input system that now also contains the environment inputs $\ket{\nu^{(k)}}_{E_k}$. We assume that the dilations are chosen such that the environment input $\ket{\nu^{(k)}}_{E_k} = \ket{\nu}_{E}$ is the same for all combs. Hence, the input to the causal structure is $\ket{\tilde{\psi}}_{\tilde{S}}= \ket{\psi}_S \otimes \ket{\nu}_E$. Since the dilation environment will be discarded only after the main protocol has finished, we can model superpositions of pure combs only and trace out the environment after the implementation (a pure comb is a sequence of unitaries with memory~\cite{TulioSuperchannels}). A schematic picture for the bipartite case is shown in Fig.~\ref{Figure:MixedCombs}.
In general, the resulting process can depend on the choice of purification. In other words, there may be several ways how to dilate and coherently control mixed combs and we consider the particular choice a part of the definition of what it means to coherently control mixed combs.

 \begin{figure}[h!]
	\begin{center}
	\includegraphics[width=0.6\textwidth]{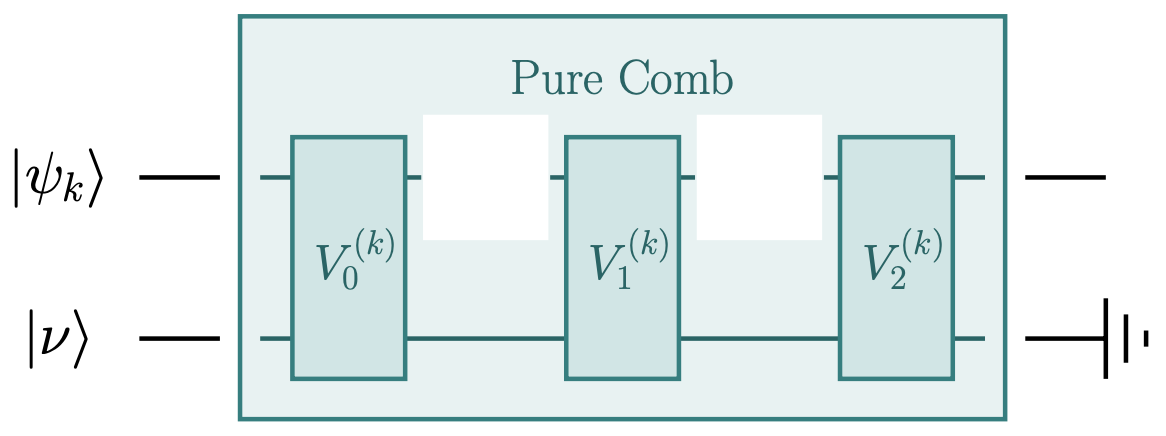}
	\end{center}
	\caption{The relation between pure and mixed combs: The dilation environment input $\ket{\nu}$ is treated as part of an extended main system that is the input to the causal structure. The partial trace over the environment output  is only applied after the main protocol has finished. The unitaries with memory are a pure comb that we handle just as in the previous sections. One can assume that the environment inputs for all combs are all the same, or that each comb has its own one and that the environment inputs of the other combs get discarded.}
	\label{Figure:MixedCombs}
\end{figure}

Each of the combs $\mathcal{\tilde G}_k$ has a definite order of the agents that we describe via a permutation $\pi_k$. More specifically, the $j$-th agent in comb $k$ is given by $A_{\pi_k(j)}$. The $V_j^{(k)}$ act trivially on the agents' ancillas $S' = A'_{1} \otimes \dots \otimes A'_N$, while the agents' unitaries $U_j$ do not act on the memory of the comb, see Fig.~\ref{Figure:CoherentlyControlled} for the tripartite example. The comb memory wire parallel to the action of agent $A_j$ is called $E_j$. As mentioned above, we assume that its dimension is independent of the comb index $k$ and, hence, the combs can be written as
\begin{align}
	\mathcal{\tilde G}_k(U_{1},U_{2},\dots U_N) = (V^{(k)}_N \otimes \mathbb 1_{S'}) (U_{{\pi_k(N)}} \ox \mathbb 1_{E_{\pi_k(N)}}) (V^{(k)}_{N-1} \ox \mathbb 1_{S'}) \dots (V^{(k)}_1 \ox \mathbb 1_{S'}) (U_{{\pi_k(1)}} \ox \mathbb 1_{E_{\pi_k(1)}}) (V^{(k)}_0 \ox \mathbb 1_{S'}).
\end{align}
Leaving identity operations on the ancillas implicit for notational convenience we thus arrive at Equation~\eqref{Eq:CohCoCausal} from the main text. \\

\begin{figure}[h!]
\includegraphics[width = 0.75\textwidth]{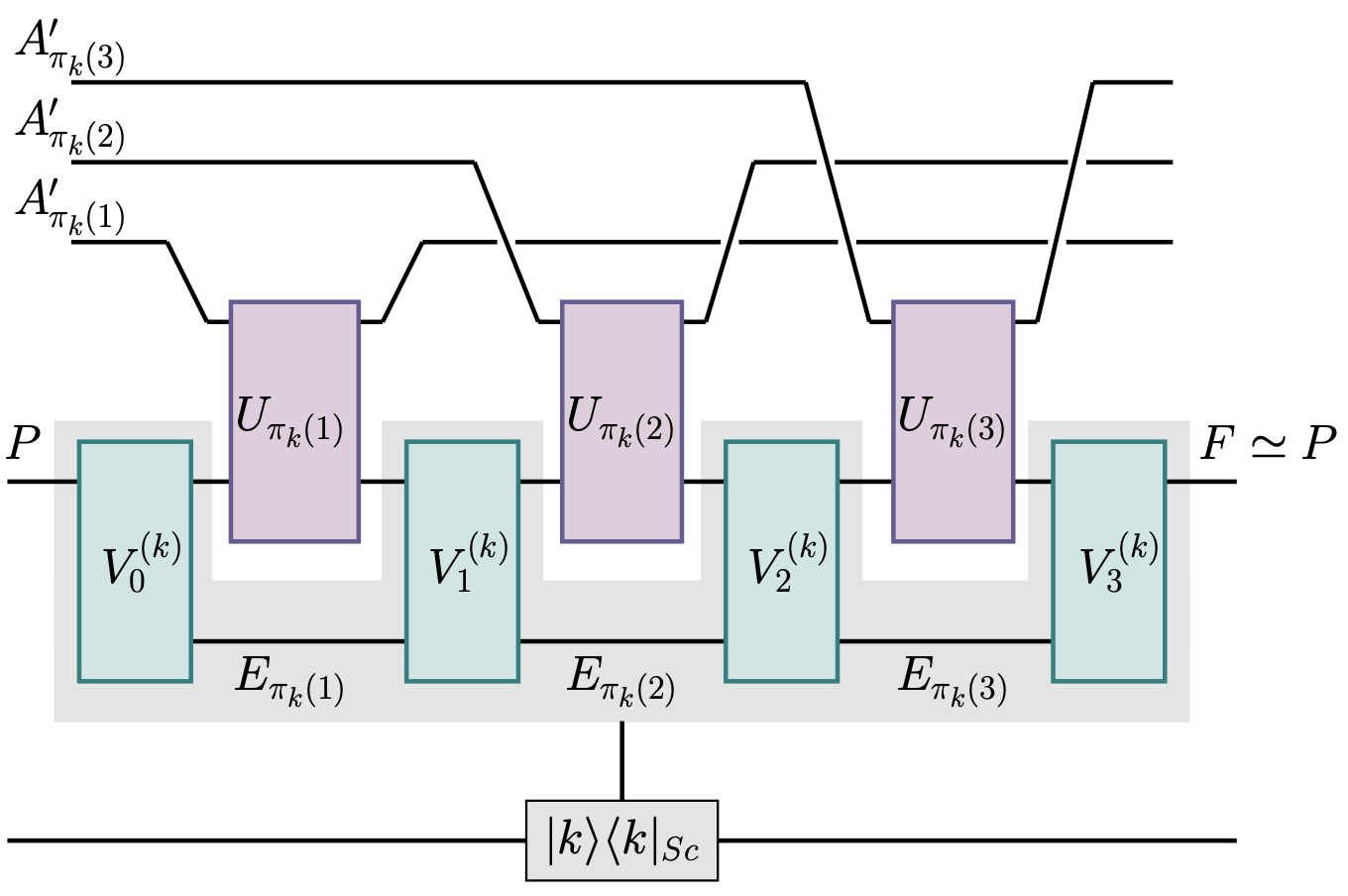}
\caption{This figure shows the general scenario for coherently controlled causal order in the tripartite case (see also \cite{PurvesQuantumControlledOrder,WechsQuantumControlledOrder}). The process $\mathcal G$ consists of a control degree of freedom, whose value $k$ controls which comb $\mathcal{\tilde G}_k$ is implemented. The order of the agents $A_j$ in comb $\mathcal{\tilde G}_k$ is described by a permutation $\pi_k$, i.e. the $m$-th agent in comb $\mathcal{\tilde G}_k$ is agent $A_{\pi_k(m)}$. The combs $\mathcal{\tilde G}_k$ are assumed to be pure, implemented via unitaries $V_j^{(k)}$ with memories and have have been extended (e.g. via ancillas) such that all the relevant dimensions are independent of $k$. This means the global past $P$ and the global future $F$ are independent of $k$. Furthermore, the dimension of the comb memory $E_{\pi_k(m)}$ running parallel to agent $A_{\pi_k(m)}$ is independent of $k$. The agents' operations $U_{j}$ are shown in purple. They can act on the ancilla $A'_{j}$ of the respective agent, but not on the ancillas of the other agents or the comb memory.}
\label{Figure:CoherentlyControlled}
\end{figure}

In what follows we describe a general procedure for writing down a history state complying with our axioms from Section~\ref{Section:OurFormalism}. While there are potentially many ways to write down such history states, our goal was to pick one with with a notation that is as simple as possible for an arbitrary number of agents. This means the procedure will not be as efficient or short as possible, but will use indices and notation that make it easier to discuss the local perspectives later on.

As written in the main text the history state decomposes into three parts as 
\begin{align}
	\ket{\Psi \rangle} = \ket{\Psi_{\text{desync}} \rangle} + \ket{\Psi_{\text{combs}} \rangle} + \ket{\Psi_{\text{resync}}\rangle},
\end{align}
and we will now consider each part separately.

\subsection{Desynchronizing the clocks}
%In the first step of the protocol we wish to manipulate the order of the agents by manipulating their clocks. 
In the first step of the protocol we use the control degree of freedom to desynchronize the clocks such that the agents are put into the right order. 
It will be helpful to manipulate the clocks such that consecutive agents are \emph{two} ticks apart because between the actions of two consecutive agents there is a unitary $V^{(k)}_j$ of the comb. To desynchronize the clocks, we will partially freeze them in time. More specifically, we start from $\ket{0,0,\dots,0}\ox \ket{\psi}_S$. At first all the clocks make two synchronized step to $ \ket{2,2,\dots,2} \otimes \ket{\psi}_S$. For the desynchronization procedure we consider a history state of the following form:
\begin{align}
	\kket{\Psi_{desync}} = \ket{0,0,\dots,0}_c \ox \ket{\psi}_S  + \ket{1,1,\dots,1}_c \ox\ket{\psi}_S + \sum_{k=1}^M\sum_{j=2}^{T_0}   \ket{t(j)^{(k)}_1, t(j)^{(k)}_2, \dots, t(j)^{(k)}_N}_c \ox (\ket{k}\bra{k}\otimes \mathbb{1}) \ket{\psi}_S 
\end{align}

There are many desynchronization procedures one can choose from. Our goal is to pick one with with a notation that is as simple as possible for an arbitrary amount of agents. This means the procedure will not be as efficient or short as possible, but will use indices and notation that make it easier to discuss the local perspectives later on. One such procedure works as follows:\\

The clock of the fastest agent, i.e. $\pi_k(1)$, continues to tick at the same rate as before. This we describe via 
\begin{align}
	t(j)_{\pi_k(1)}^{(k)} = j
\end{align}
For notational simplicity, we will desynchronize the clocks one after the other. Consider integers $2 \le m \le N$. We use the time range described by $2(m-2)\cdot N +2 \le j \le 2(m-1)\cdot N+1$ to slow down the clock of agent $\pi_k(m)$. More specifically, at times $2(m-2)\cdot N + 2 \le j \le 2(m-2)\cdot N + 2(m-1)+2$, the clock of agent $m$ completely freezes, while the clocks of the other agents march on. Except for that freezing period, the clock ticks at a normal rate. Overall, this can be described as follows ($m\ge 2$):
\begin{align}
	t(j)_{\pi_k(m)}^{(k)} = 
	\begin{cases}
		j  & \text{ for } j \le 2(m-2)N+2 \\
		2(m-2)\cdot N +2 & \text{ for } 2(m-2)N+2 \le j \le 2(m-2)N +2(m-1)+2 \\
		j - 2(m-1) & \text{ for } j \ge 2(m-2)N +2(m-1)+3
	\end{cases} \label{Equation:t}
\end{align}
We choose the largest $j$ to be 
\begin{align}
	T_0 := 2(N-2)N +2(N-1)+4+2(N+1) = 2N^2+4,
\end{align}
which includes $2(N+1)$ more well-synchronized ticks to make sure that for all $k$ the clocks freezes are far way from the application of the combs. The desynchronization procedure is shown for $N=4$ in Figure \ref{Figure:Times}.
 
 \begin{figure}[h!]
	\begin{center}
	\includegraphics[width=\textwidth]{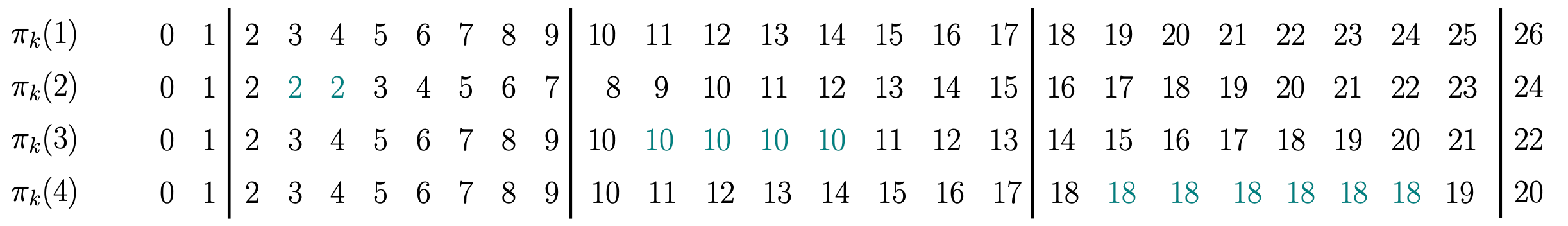}
	\end{center}
	\caption{This figure shows the clock times during the desynchronization procedure for the special case $N=4$. Time passes from left to right. Time freezing is marked in color. After the shown times only well-synchronized ticks happen.}
	\label{Figure:Times}
\end{figure}

First, we note that two consecutive agents $\pi_{k}(m)$ and $\pi_k(m+1)$ are indeed two time steps apart at the end: 
\begin{align*}
 	[j - 2([m+1]-1)] - [ j- 2(m-1)] = -2
\end{align*}
Moreover, we made sure to construct the history state such that only one clock freezes simultaneously, i.e. time freezes of different clocks are well-separated from each other and that no times are skipped.

Next, let us consider the local perspectives, i.e. $_{c_a}\braket{t}{\Psi_{desync}\rangle}$.
Let us expand $\ket{\psi}_S = \sum_{k} \ket{k}_{Sc} \ket{\psi_k}_{Sp}$. Then we have 
\begin{align}
	\ket{\Psi_{desync} \rangle} =  \ket{0,0,\dots,0}_c \ox \ket{\psi}_S +  \ket{1,1,\dots,1}_c\ox \ket{\psi}_S  + \sum_{k=1}^M  \sum_{j=2}^{T_0}  \ket{t(j)^{(k)}_1, t(j)^{(k)}_2, \dots, t(j)^{(k)}_N}_c \ox \ket{k}_{Sc} \ket{\psi_k}_{Sp} 
\end{align}

%From $t_a=1$ on, the time-evolution agent $A_a$ will see controlled unitaries of the form $\sum_{k=1}^M \ket{k}\bra{k}\otimes U_{k}^{(a)}(t_a+1,t_a)$. Explicitly writing it down is very convoluted, so we will describe it instead. Usually, one agent just sees the clocks of the other agents perform a tick, as modeled via $T$ with $\ket{j}\mapsto\ket{j+1}$. There are two cases that are different. One case is if the clock of the considered agent freezes. The other case is if the clock of another agent freezes. 

{We now need to define the normalisation operators $N_t^{A_j}$ and the unitaries $\UU_{A_j} (t, t')$ relating the perspectival states at different times. Without loss of generality, we will show how to construct those for the point of view of $A_1$. Define 
\begin{equation}
\alpha_k(t) = \lVert \bra{ t}_{c_1} \sum_{j=2}^{T_0}  \ket{t(j)^{(k)}_1, t(j)^{(k)}_2, \dots, t(j)^{(k)}_N}_c \rVert.
\end{equation}
Note that for any $1<t < T_{A_1}$, where $T_{A_1}$ is the largest time $A_1$ sees during the desynchronization phase, we have that $\alpha_k(t) \neq 0$ because no time is skipped during desynchronization. We can therefore define
\begin{equation}
N_t^{(A_1)} = \sum_k \frac{1}{\alpha_k(t) } \ket{k} \bra{k}_{S_c} ,
\end{equation}
as the normalization operator, which then gives the perspectival state
\begin{equation}
\ket{\psi^{A_1}(t)} = \sum_k \ket{\xi_k}_{c} \ket{k}_{S_c} \ket{\psi_k}_{S_p},
\end{equation}
where $\ket{\xi_k(t)}_c$ is a normalized state proportional to $  \bra{t}_{c_1} \sum_{j=2}^{T_0}  \ket{t(j)^{(k)}_1, t(j)^{(k)}_2, \dots, t(j)^{(k)}_N}_c$. It is clear that there exists a unitary relating $\ket{\psi^A_1(t)}$ with $\ket{\psi^A_1(t+1)}$ and that this unitary can be chosen to have the form $\UU_{A_1}(t, t+1) = \sum_k  u^{A_1}_{c,k}\ox \ket{k}\bra{k}_{S_c}\otimes \id_{S_p} $. Indeed, we can choose $u^{A_1}_{c,k}$ to be any unitary mapping $\ket{\xi_k (t)}_c \mapsto \ket{\xi_k(t+1)}$, and acting arbitrarily on other states. 
}

\subsection{Application of the combs}
Now we consider the application of the combs. The starting point is 
\begin{align}
\sum_{k=1}^M (\ket{k}\bra{k}\otimes \mathbb{1}) \ket{\psi}_S \otimes  \ket{T_0, T_0-2, \dots, T_0 - 2(N-1)}_{c_{\pi_k(1)}, \dots, c_{\pi_k(N)}},
\end{align}
with
\begin{equation}
\ket{t_1, t_2, \dots, t_N}_{c_{\pi_k(1)}, \dots, c_{\pi_k(N)}}:=U_{\pi_k} \ket{t_1, t_2, \dots, t_N}_{c},
\end{equation}
where $U_{\pi_k}$ is the unitary implementing the permutation on the Hilbert spaces of the local clocks.

For this part of the protocol, the clocks will always tick in synchronization. Then all agents see the following sequence of time evolutions:
\begin{align}
	V^{(k)}_0 \otimes T^{\otimes (N-1)},\ U_{{\pi_k(1)}}\otimes T^{\otimes (N-1)},\  V^{(k)}_1\otimes T^{\otimes (N-1)}, \ U_{{\pi_k(2)}}\otimes T^{\otimes (N-1)}, \ \dots, \ U_{{\pi_k(N)}}\otimes T^{\otimes (N-1)}, \ V^{(k)}_N \otimes T^{\otimes (N-1)}
\label{eq:perspectives_for_combs} 
\end{align}
So the time of action for each agent is $t^* = T_0 + 2$. For completeness, let us describe the time-evolutions the agents see in more detail. For that purpose, we start at $\tau := T_0 -2(N-1)$. Then the unitary time evolution that agent $j$ sees are given by ($p$ a non-negative integer)
\begin{align}
	\UU_{A_j}(\tau + p + 1, \tau +p) = \sum_{k=1}^M \proj{k}_{Sc} \otimes T^{\otimes (N-1)} \otimes W_{A_j}^{(k)}(p+1,p)
\end{align} 
Let $m_j^{(k)}$ be the integer with $\pi_k(N-m_j^{(k)}) = j$. Then the unitary $W_{A_j}^{(k)}(p+1,p)$ is given by ($ N \ge x\ge 0$ a non-negative integer, $N \ge y \ge 1$ a positive integer)
\begin{align}
	&W_{A_j}^{(k)}(2m_j^{(k)} + 2x +1, 2m_j^{(k)} + 2x) = V_x^{(k)}, \nonumber \\
	&W_{A_j}^{(k)}(2m_j^{(k)} + 2y, 2m_j^{(k)} + 2y-1) = U_{\pi_k(y)}, \\
	&W_{A_j}^{(k)}(p+1, p) = \mathbb 1 \text{ for other  values of } p. \nonumber
\end{align}

The corresponding part of the history state looks as follows 
\begin{align}
\ket{\Psi_{combs} \rangle} =
& {\sum_{k=1}^M \ket{T_0 +1, T_0-1 , \dots, T_0 -2(N-1)+1}_{c_{\pi_k(1)},\dots c_{\pi_k(N)}} \ox \Big[\ket{k}\bra{k}\otimes V^{(k)}_0\Big] \ket{\psi}_S }+ \nonumber\\
&+\sum_{k=1}^M \sum_{y=1}^{N} \ket{T_0 +2y, T_0 -2 +2y, \dots, T_0 -2(N-1)+2y}_{c_{\pi_k(1)},\dots c_{\pi_k(N)}} \nonumber \\
& \qquad \qquad  \qquad \qquad \qquad  \qquad \ox \Big[\ket{k}\bra{k}\otimes \Big( U_{\pi_{k}(y)} V^{(k)}_{y-1} \dots U_{{\pi_k(1)}} V^{(k)}_0\Big) \Big] \ket{\psi}_S 
+ \nonumber \\
&+\sum_{k=1}^M \sum_{x=1}^{N}  \ket{T_0 +1 +2x, T_0-1 +2x, \dots, T_0 -2(N-1)+1+2x}_{c_{\pi_k(1)},\dots c_{\pi_k(N)}}\nonumber  \\ 
& \qquad \qquad  \qquad \qquad \qquad  \qquad \ox \Big[\ket{k}\bra{k}\otimes \Big( V^{(k)}_{x} U_{\pi_{k}(x)} \dots U_{{\pi_k(1)}} V^{(k)}_0\Big) \Big] \ket{\psi}_S  \label{eq:all_combs}
\end{align}  

Hence, all the combs get applied, see Eq.~\eqref{eq:all_combs}, each agent has a well defined time of action and the other agent's unitaries appear at most linearly in each parties perspective, see Eq.~\eqref{eq:perspectives_for_combs}.

%%%%%%%%%%%%%%%%%
\subsection{Resynchronization}
Now we consider the final part of the protocol, the resynchronization step. We define 
\begin{align}
	T_{1} := T_0 +2N+1
\end{align} 
such that the starting point is given by 
\begin{align}
	&\sum_{k=1}^M \ket{T_1, T_1-2, \dots, T_1 -2(N-1)}_{c_{\pi_k(1)},\dots c_{\pi_k(N)}} \ox \Big[\ket{k}\bra{k}\otimes \Big( V^{(k)}_{N} U_{\pi_{k}(N)} \dots U_{A_{\pi_k(1)}} V^{(k)}_0\Big) \Big] \ket{\psi}_S \nonumber
\end{align}
To make sure that for all $k$ the clock freezes are far apart from the application of the combs, we first insert $2(N+1)$ well-synchronized ticks. Afterwards, we choose the resynchronization to proceed exactly as the desynchronization, but with the order of agents reversed. By using the function $t(j)^{(k)}_{\pi_k(m)}$ from Eq.~\eqref{Equation:t}, this can be described by the history state

\begin{align}
	\kket{\Psi_{resync}} &= \nonumber  \\
	  & \sum_{k=1}^M \sum_{j=0}^{2(N+1)} \ket{T_1 +1+ j, T_1-1 +j, \ \dots,\ T_1+1 -2(N-1)+ j}_{c_{\pi_k(1)},\ \dots c_{\pi_k(N)}}    \nonumber \\
	  & \qquad \qquad  \qquad \qquad \qquad  \qquad \ox  \Big[\ket{k}\bra{k}\otimes \Big( V^{(k)}_{N} U_{\pi_{k}(N)} \dots U_{{\pi_k(1)}} V^{(k)}_0\Big) \Big] \ket{\psi}_S +\\
	&+\sum_{k=1}^M \sum_{j=0}^{T_0} \ket{T_1 +2(N+1) +2+ t(j)^{(k)}_{\pi_k(N)},\ T_1+2(N+1) +  t(j)^{(k)}_{\pi_k(N-1)}, \ \dots,\ T_1+6+ t(j)^{(k)}_{\pi_k(1)}}_{c_{\pi_k(1)},\ \dots c_{\pi_k(N)}} \nonumber  \\ 
	 & \qquad \qquad  \qquad \qquad \qquad  \qquad \ox   \Big[\ket{k}\bra{k}\otimes \Big( V^{(k)}_{N} U_{\pi_{k}(N)} \dots U_{{\pi_k(1)}} V^{(k)}_0\Big) \Big] \ket{\psi}_S  \nonumber 
\end{align}

Just as during the desynchronization process, nothing happens on the system and we can write the perspectival states and unitaries as 
$\ket{\psi^{A_j}(t)} = \sum_k \ket{\xi_k}_{c}  \mathcal{\tilde G}_k(U_{1},U_{2},\dots U_N)\ket{k}_{S_c} \ket{\psi_k}_{S_p}$ and 
$\mathcal{U}_{A_j}(t, t+1) = \sum_k  V_{c}^k \ox \ket{k}\bra{k}_{S_c}\otimes \id_{S_p} $. \\

With this generic protocol we can implement any process describing the coherent control of causal order within our Page-Wootters framework.

%%%%%%%%%%%%%%%%%%%%%%%%%%%%%%%%%%%%%%%%%%%%%%%%%%%%%%%

\end{document}